\documentclass{emulateapj}
\usepackage{natbib,aas_macros}
\usepackage{multirow,color}

\begin{document}
\shorttitle{Demographics of Lyman-Alpha Emission in $z \sim 7$ Galaxies}
\shortauthors{Ono et al.}
\slugcomment{Accepted for publication in ApJ}

\title{%
Spectroscopic Confirmation of Three $\lowercase{z}$-Dropout Galaxies at $\lowercase{z} = 6.844 - 7.213$: \\
Demographics of Lyman-Alpha Emission in $\lowercase{z} \sim 7$ Galaxies\altaffilmark{\ddag}
}
\author{%
Yoshiaki Ono~\altaffilmark{1},
Masami Ouchi~\altaffilmark{2,3},
Bahram Mobasher~\altaffilmark{4},
Mark Dickinson~\altaffilmark{5}, 
Kyle Penner~\altaffilmark{6},\\
Kazuhiro Shimasaku~\altaffilmark{1,7}, 
Benjamin J. Weiner~\altaffilmark{8},  
Jeyhan S. Kartaltepe~\altaffilmark{5},
Kimihiko Nakajima~\altaffilmark{1}, \\
Hooshang Nayyeri~\altaffilmark{4}, 
Daniel Stern~\altaffilmark{9}, 
Nobunari Kashikawa~\altaffilmark{10},
and Hyron Spinrad~\altaffilmark{11}
}

\email{ono \_at\_ astron.s.u-tokyo.ac.jp}

\altaffiltext{1}{%
Department of Astronomy, Graduate School of Science,
The University of Tokyo, Tokyo 113-0033, Japan
}
\altaffiltext{2}{%
Institute for Cosmic Ray Research, University of Tokyo,
Kashiwa 277-8582, Japan
}
\altaffiltext{3}{%
Institute for the Physics and Mathematics of the Universe (IPMU), TODIAS,
University of Tokyo, 5-1-5 Kashiwanoha, Kashiwa, Chiba 277-8583, Japan
}
\altaffiltext{4}{%
Department of Physics and Astronomy, University of California, 
Riverside, CA, 92521, USA
}
\altaffiltext{5}{%
National Optical Astronomical Observatories, 
Tucson, AZ 85719, USA
}
\altaffiltext{6}{%
Department of Astronomy, University of Arizona, 
933 N. Cherry Ave., Tucson, AZ 85721, USA
}
\altaffiltext{7}{%
Research Center for the Early Universe, Graduate School of Science,
The University of Tokyo, Tokyo 113-0033, Japan
}
\altaffiltext{8}{%
Steward Observatory, University of Arizona, 
933 N. Cherry Avenue, Tucson, AZ 85721, USA
}
\altaffiltext{9}{%
Jet Propulsion Laboratory, California Institute of Technology, 
4800 Oak Grove Drive, Pasadena, CA 91109, USA
}
\altaffiltext{10}{%
Optical and Infrared Astronomy Division, National Astronomical Observatory of Japan, 
2-21-1, Osawa, Mitaka, Tokyo, 181-8588, Japan
}
\altaffiltext{11}{%
Department of Astronomy, University of California, Berkeley, 
CA 94720, USA
}
\altaffiltext{\ddag}{%
Based on data obtained with the Subaru Telescope and the W. M. Keck
Observatory. The Subaru Telescope is operated by the National Astronomical
Observatory of Japan. The W. M. Keck Observatory is operated as a scientific
partnership among the California Institute of Technology, the University of
California, and the National Aeronautics and Space Administration.
}

\begin{abstract}
We present the results of our ultra-deep Keck/DEIMOS spectroscopy
of $z$-dropout galaxies in the SDF and GOODS-N.
For $3$ out of $11$ objects, 
we detect an emission line 
at $\sim 1\mu$m with a signal-to-noise ratio of $\sim 10$. 
The lines show asymmetric profiles with high weighted skewness values,  
consistent with being Ly$\alpha$, 
yielding redshifts of $z=7.213$, $6.965$, and $6.844$.
Specifically, we confirm the $z=7.213$ object 
in two independent DEIMOS runs 
with different spectroscopic configurations.
The $z=6.965$ object is a known Ly$\alpha$ emitter, IOK-1, 
for which our improved spectrum 
at a higher resolution yields a robust skewness measurement. 
The three $z$-dropouts have Ly$\alpha$ fluxes of 
$3 \times 10^{-17}$ erg s$^{-1}$ cm$^{-2}$
and rest-frame equivalent widths EW$_0^{{\rm Ly}\alpha} = 33-43${\AA}.
Based on the largest 
spectroscopic sample of $43$ $z$-dropouts
that is the combination of our and previous data, we find
that the fraction of Ly$\alpha$-emitting galaxies
(EW$_0^{{\rm Ly}\alpha} > 25${\AA}) is low at $z \sim 7$;
$17 \pm 10${\%} and $24 \pm 12${\%} 
for bright ($M_{\rm UV} \simeq -21$) and 
faint ($M_{\rm UV} \simeq -19.5$) galaxies, respectively.
The fractions of Ly$\alpha$-emitting galaxies
drop from $z \sim 6$ to $7$ 
and the amplitude of the drop
is larger for faint galaxies than for bright galaxies.
These two pieces of evidence would indicate
that the neutral hydrogen fraction of the IGM increases
from $z \sim 6$ to $7$, 
and that the reionization proceeds 
from high- to low-density environments,
as suggested by an inside-out reionization model.
\end{abstract}


\section{INTRODUCTION} \label{sec:introduction}

Over the last two years, 
we have witnessed an explosion of activities 
aimed at searching for very high redshift ($z > 7$) galaxies. 
This has been made possible by the installation 
of the Wide-Field Camera 3 (WFC3) 
on-board the Hubble Space Telescope \citep[e.g.,][]{oesch2010,oesch2011,bouwens2010,bouwens2011,bouwens2011b,mclure2009,mclure2011,wilkins2010,wilkins2011,bunker2010,yan2010,yan2010b,lorenzoni2010},  
the Hawk-I instrument on the Very Large Telescope \citep[VLT;][]{castellano2010,castellano2010b}, 
and improvement in the red-end sensitivity of 
the Suprime-Cam on the Subaru telescope \citep{ouchi2009b}. 
These studies have identified over a hundred candidates at $z > 6.5$ 
through the Lyman break drop-out technique \citep[e.g.,][]{steidel1996a,giavalisco2002},  
showing a decrease in the number density of 
bright high-redshift galaxies 
with redshift \citep[e.g.,][]{ouchi2009b,mclure2009,castellano2010,bouwens2010,wilkins2010}, 
bluer UV continua \citep[e.g.,][]{bouwens2010b}\footnote{See also 
\cite{dunlop2011}, who question this result.}, 
and relatively smaller stellar masses compared to lower redshift galaxies 
selected by similar techniques \citep[e.g.,][]{labbe2010,schaerer2010,finkelstein2009f,ono2010}. 
Study of the intrinsic properties of high-redshift galaxies at epochs 
close to the dark ages is essential 
for understanding one of the most outstanding questions 
in modern astronomy - 
when did the Universe become reionized 
and what sources were responsible for it?
This can be accomplished by studying the state of 
the inter-galactic medium (IGM) through estimates 
of the ionizing photon budget and Lyman $\alpha$ escape fraction.

Several independent studies of the reionization process in recent years 
have yielded different results. 
By measuring the polarization of the cosmic background radiation, 
\cite{dunkley2009} estimated the optical depth to reionization 
and concluded that, if it was a sudden event, 
reionization occurred at $z = 11.0 \pm 1.4$  \citep[see also,][]{komatsu2010,larson2011}. 
However, investigating the spectra of SDSS quasars, 
\cite {fan2006} studied the evolution of the Gunn-Peterson optical depth 
and demonstrated that the IGM reionization 
may have ended as late as $z \sim 6$. 
The result from this study is questioned by \cite{goto2011c}, 
who have shown that for quasars at $z > 6$, 
a statistically large number is needed 
in $\Delta z = 0.1$ bins to constrain cosmic variance 
and trace the evolution of the optical depth.

Another useful tool for studying reionization is 
the Lyman $\alpha$ luminosity function of 
high-redshift galaxies 
selected by narrow-band imaging \citep[e.g.,][]{hu1998,rhoads2000}; 
these galaxies are called Lyman Alpha Emitters (LAEs). 
Since neutral hydrogen in the IGM resonantly scatters Ly$\alpha$ photons, 
the transmission of Ly$\alpha$ is sensitive 
to the ionization state of the IGM. 
Therefore, we expect a decrease in the number density of LAEs 
close to the reionization epoch \citep[e.g.,][]{haiman1999,malhotra2004,santos2004,mesinger2004,stern2005,haiman2005,furlanetto2006,mcquinn2007,dijkstra2007,kobayashi2007,mesinger2008b,iliev2008,dayal2008,dayal2009,dayal2011}. 
\cite{ouchi2010} found a decrease 
in the Ly$\alpha$ luminosity function of LAEs, 
corresponding to an upper limit of 
the IGM neutral fraction $x_{\rm HI} \lesssim 0.2 \pm 0.2$ at $z=6.6$, 
which indicates that the major reionization process 
took place at higher redshift. 
This result has been further supported by \cite{kashikawa2011}, 
who studied the evolution of the LAE luminosity function 
in the Subaru Deep Field \citep[SDF;][]{kashikawa2004} 
and found an increase in the neutral fraction of the IGM 
from $z=5.7$ to $6.5$. 
\cite{nakamura2011} have also found a large deficit in the number density 
of $z=6.5$ LAEs in the SSA22 field, 
and attributed it to significant field-to-field variance 
of the neutral fraction of the IGM. 
The effect of cosmic variance was quantified by \cite{ouchi2010}, 
who claimed a factor of $2-10$ range in the number density of LAEs 
in an extensive survey covering an area of $1$ deg$^2$.

\begin{deluxetable*}{cccccccc} 
\tablecolumns{8} 
\tablewidth{0pt} 
\tablecaption{Summary of observations with Keck/DEIMOS.\label{tab:sum_obs}}
\tablehead{
    \colhead{Mask ID}    & \colhead{Field} &  \colhead{Date (UT)} 
    & \colhead{Total Exposure}  & \colhead{$N_z$\tablenotemark{$\dagger$}}    & \colhead{grating}   & \colhead{central wavelength}   & \colhead{filter} \\
    \colhead{ } & \colhead{ } & \colhead{ } 
    & \colhead{[sec]} & \colhead{ } & \colhead{[lines mm$^{-1}$]} & \colhead{[{\AA}]} &  \colhead{ } 
}
\startdata 
SDFZD1B & SDF  &   2010 February 13  & $19350$  &  $5$  &  $830$  &  $9000$  &  OG550  \\
SDFZD3 & SDF  &   2010 April $14-15$ & $30000$  &  $2$  &  $830$  &  $9000$  &  OG550  \\
SDFZD4 & SDF  &   2010 April 15 & $7200$  &  $4$  &  $830$  &  $9000$  &  OG550  \\
GNZD1B & GOODS-N  &  2010 February 13, April 14-15 & $18000$  &  $2$  &  $830$ &  $9000$  &  OG550  \\
HDF11C & GOODS-N  &  2011 April $1-2$ & $14600$  &  $2$  &  $600$ &  $7500$  &  GG455  \\
HDF11D & GOODS-N  &   2011 April 3 & $7200$  &  $2$  &  $830$ &  $8100$  &  OG550  
\enddata 
\tablenotetext{$\dagger$}{%
Numbers of observed $z$-dropouts. 
Some objects were observed on multiple masks (See Table \ref{tab:sum_followup}). 
}
\end{deluxetable*} 

\begin{figure*}
\hspace{0.3cm}
   \includegraphics[scale=0.75]{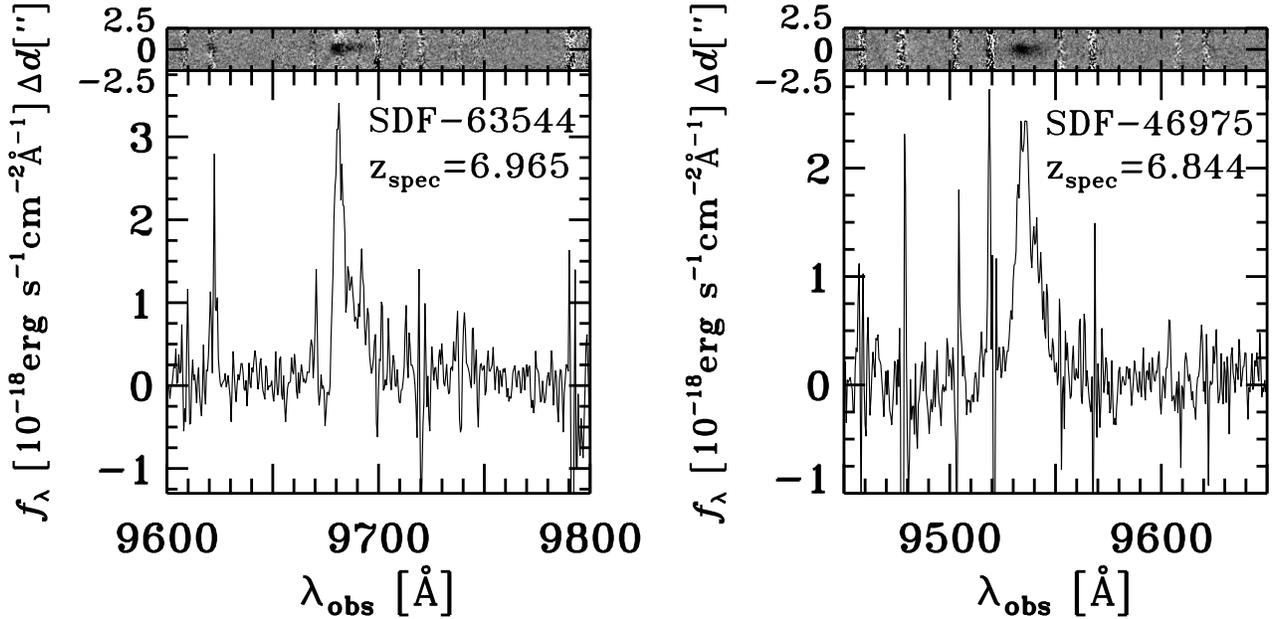}
 \caption[]
{
Spectra of SDF-63544 (left) and SDF-46975 (right). 
The top panels show the composite two-dimensional spectra, 
from which the one-dimensional spectra 
shown in the bottom panels are derived. 
A prominent emission line is seen 
at $9683$ {\AA} for SDF-63544 ($S/N \simeq 13$), 
and 
at $9536$ {\AA} for SDF-46975 ($S/N \simeq 14$). 
}
\label{fig:2d1dspectra}
\end{figure*}

Measuring the fraction of LAEs 
among Lyman-break Galaxies \citep[LBGs;][]{giavalisco2002}, 
the Ly$\alpha$ fraction, 
provides complementary information 
to understand the reionization process \citep[e.g.,][]{stark2010}. 
Since LBGs are selected over a broader range of redshifts 
compared to observations with a narrow-band filter, 
their number density is less sensitive to cosmic variance. 
Searching for Ly$\alpha$ emission from samples of LBGs 
with available spectra at $4 < z < 6$, 
\cite{stark2010b} showed that lower luminosity LBGs have 
larger Ly$\alpha$ fractions, 
and that the fraction increases with redshift \citep[see also,][]{vanzella2009,stark2010,douglas2010}. 
However, it is not yet clear if this trend continues $z > 6$. 
To explore this, we require spectroscopy of dropout candidates at $z \sim 7$.

The real challenge in estimating the number density of $z > 6$ galaxies 
and the intensity of ionizing Ly$\alpha$ photons 
is the spectroscopic confirmation of these candidates. 
They are extremely faint, and the only detectable feature 
is Ly$\alpha$ emission shifted to 
near-infrared wavelengths \citep[e.g.,][]{iye2006,lehnert2010,fontana2010,vanzella2010d}. 
Despite significant efforts 
to spectroscopically confirm $z > 6$ candidates, 
the number of confirmed sources is still very limited. 
\cite{fontana2010} reported the detection of 
one LBG with Ly$\alpha$ emission at $z=6.97$
in ultra-deep spectroscopy of seven $z$-dropout candidates 
selected from VLT/Hawk-I imaging 
of the Great Observatories Origins Deep Survey's southern 
(GOODS-S) field \citep{castellano2010}. 
They found a significant decline in the fraction of LBGs with Ly$\alpha$ emission 
between $z \sim 6$ and $7$, 
reversing the increasing trend with redshift found at $z<6$. 
Furthermore, \cite{vanzella2010d} spectroscopically confirmed 
two $z$-dropout galaxies at $z \approx 7.1$ 
selected from VLT/Hawk-I imaging 
of the BDF4 field \citep{castellano2010b}. 
With the small number of $z$-dropout galaxies 
with available spectroscopic redshifts, 
any measure of their Ly$\alpha$ photon budget or number density
will be seriously affected 
by statistical uncertainties and cosmic variance.

In this study 
we present results from ultra-deep spectroscopy with Keck/DEIMOS 
for a sample of $11$ $z$-dropout galaxies. 
To maximize the spectroscopic success, 
we designed the photometric survey to identify dropout candidates 
at the bright end of the UV luminosity function. 
The aim is to derive the fraction of LBGs with Ly$\alpha$ emission 
and to study its evolution to $z \sim 7$.

In the next section 
we present the photometric selection of $z$-dropout candidates 
which are the targets for our spectroscopic observations here. 
The spectroscopic observations are described in Section \ref{sec:observations}. 
This is followed by redshift identification and the measurement of 
spectroscopic properties in Section \ref{sec:results}. 
In Section \ref{sec:discussion} 
we discuss the implications for reionization. 
The conclusions are presented in Section \ref{sec:conclusion}. 
Throughout this paper, we use magnitudes in the AB system \citep{oke1983}
and assume a flat universe 
with ($\Omega_{\rm m}$, $\Omega_{\rm \Lambda}$, $h$) $= (0.3, \, 0.7, \, 0.7)$.

\section{Photometric Selection of $\lowercase{z}$-Dropout Candidates} \label{sec:photo_selection}

In order to achieve successful spectroscopy of high redshift LBGs, 
we need to identify a sample of bright candidates. 
However, the bright end of the luminosity function 
exponentially decreases. 
Therefore, we need to cover a large area 
to find a sufficient number of LBGs 
bright enough for spectroscopy.

We carried out a wide-area photometric survey 
aimed at identifying $z$-dropout galaxies (i.e., galaxy candidates at $z \sim 7$) 
using Suprime-Cam on the Subaru telescope, 
outfitted with a custom-made filter ($y$-band) with 
effective wavelength at $1 \mu$m \citep{ouchi2009b}. 
This combination is ideal for identifying the bright population 
of LBGs at high redshifts. 
We covered an area of $1568$ arcmin$^2$ to $y \simeq 26.0$ mag 
($4 \sigma$ limit)
for two fields: 
the SDF 
and the GOODS northern field \citep[GOODS-N;][]{giavalisco2004}.  
We identified $22$ $z$-dropout candidates with $y = 25.4-26.1$ mag, 
i.e., $M_{\rm UV} < -21$.
This includes a galaxy which was already identified 
to be at a spectroscopic redshift of $6.96$ \citep{iye2006}. 
These provide the targets for the spectroscopic observations 
in this paper.

\begin{figure}
\hspace{0.1cm}
   \includegraphics[scale=0.75]{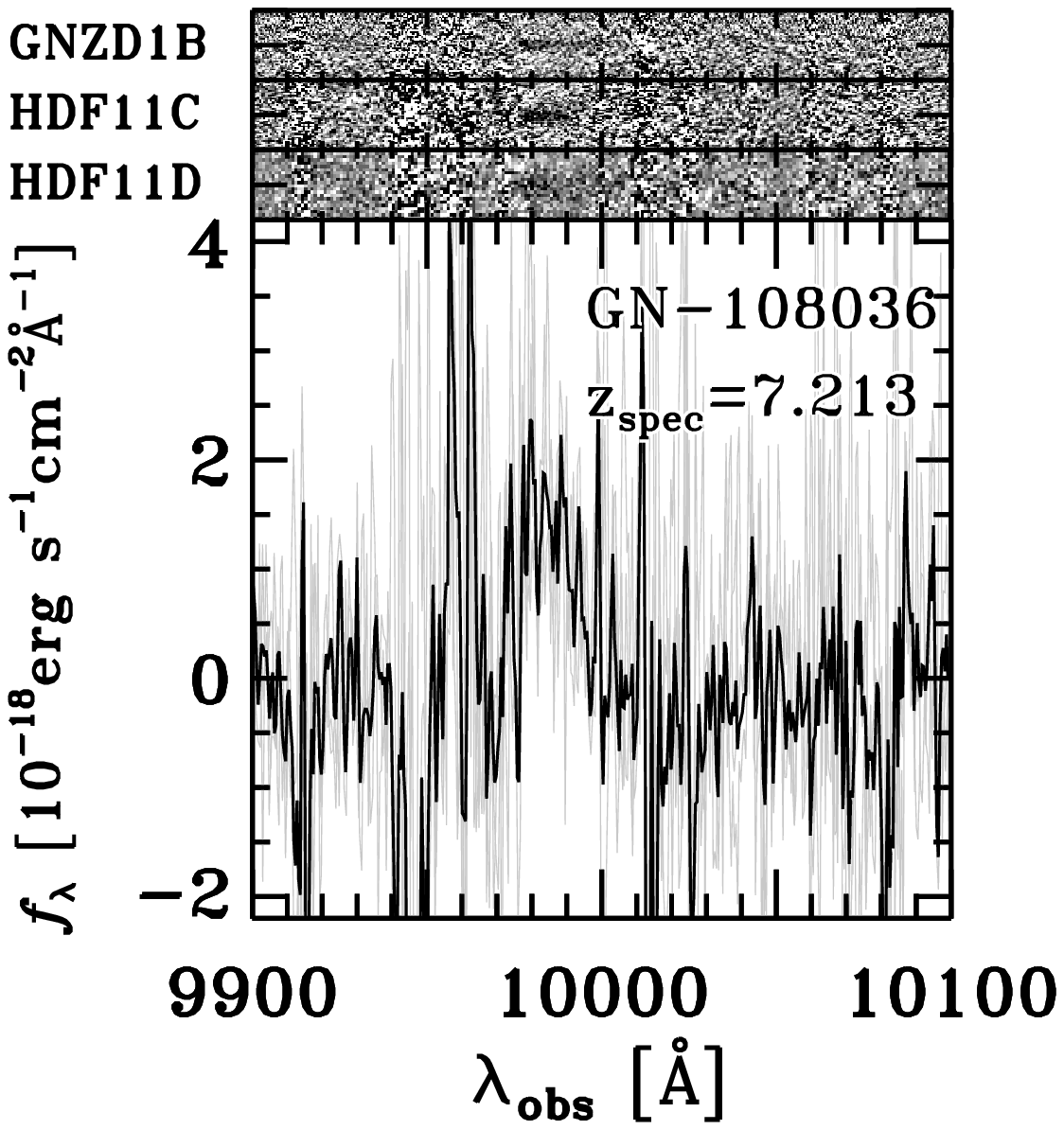}
    \caption[]
{
Spectrum of the $z=7.213$ $z$-dropout galaxy, GN-108036. 
The top panels show its two-dimensional spectra 
obtained with the GNZD1B, HDF11C, and HDF11D masks. 
The size along the spatial axis is 5.0{\arcsec} 
for each two-dimensional spectrum.
The HDF11D spectrum is binned in $2 \times 2$ pixels. 
A line is visually identified 
at $\simeq 9980${\AA} in the spectra of GNZD1B and HDF11C, 
whose exposure times are $5$ hours and $\simeq 4$ hours, respectively, 
while the line is marginally seen in the spectrum of HDF11D, 
whose exposure time is $2$ hours.
In the bottom panel, we show the one-dimensional spectra. 
The gray solid lines are 
spectra obtained with individual masks. 
The composite spectrum is shown as the black solid line. 
All the one-dimensional spectra 
illustrate a line detection at around $9980$ {\AA}, 
and the $S/N$ ratio of the line in the composite spectrum is $\simeq 6$. 
}
\label{fig:2d1dspectra_GN108036}
\end{figure}

\section{Spectroscopic Observations} \label{sec:observations}

We used the 
DEep Imaging Multi-Object Spectrograph \citep[DEIMOS;][]{faber2003} 
at the Nasmyth focus of the $10$ m Keck II telescope 
to perform spectroscopic observations of the $z$-dropout candidates 
discovered in \cite{ouchi2009b}. 
The data were taken on UT 2010 February $13$, April $14-15$, 
and 2011 April $1-3$. 
We observed $11$ out of the $22$ $z$-dropout candidates. 
We also observed the standard stars G191B2B and Wolf 1346 for flux calibration. 
The seeing was in the range $0.5'' - 0.7''$.
We used a total of six DEIMOS masks, as listed in Table \ref{tab:sum_obs}. 
For five masks, 
we used the OG550 filter and the $830$ lines mm$^{-1}$ grating, 
which is blazed at $8640${\AA} 
and was tilted to place a central wavelength of $9000$ {\AA} on the detectors. 
This configuration provided a spectral coverage between $7000$ {\AA} and $10400$ {\AA}.
For the remaining mask (HDF11C) 
we used the GG455 filter
and the $600$ lines mm$^{-1}$ grating, 
blazed at $7400${\AA}. 
This was tilted to place a central wavelength of 
$7500${\AA} on the detector. 
The spectral coverage was between 5200 {\AA} and 10200 {\AA}.
The spatial pixel scale was $0.1185{\arcsec}$ pix$^{-1}$,  
and 
the spectral dispersion was  
$0.47$ {\AA} pix$^{-1}$ 
and $0.65$ {\AA} pix$^{-1}$ 
for the $830$ lines mm$^{-1}$ 
and $600$ lines mm$^{-1}$ grating, respectively. 
The slit widths were $1''$. 
For objects filling the slit, 
the FWHM resolution of the $830$ and $600$ grating 
was $\simeq 3.3${\AA} and $\simeq  4.7${\AA}, 
respectively\footnote{\texttt{http://www2.keck.hawaii.edu/inst/deimos/specs.html}}. 
Details of the observations, 
the filters, gratings and the total exposure time used 
are listed in Table \ref{tab:sum_obs}.

We used the GG455 filter for HDF11C, 
although this filter allows transmission of second-order light 
redward of $\sim 9100$ {\AA}.  
This filter was used because 
some of the targets on those masks were 
$z \sim 3$ ultra-luminous infrared galaxy candidates, 
which would have spectral signatures below $5000$ {\AA}. 
Although this configuration allowed 
contamination from second-order light of an emission line
whose wavelength is longer than $\simeq 4500${\AA},
the targeted $z$-dropouts in the HDF11C mask 
were also observed in other configurations 
without such contaminations in the red wavelength range.

The reduction was performed using the \texttt{spec2d} IDL pipeline\footnote{The pipeline was developed 
at UC Berkeley with support from NSF grant AST-0071048. 
Downloaded at \texttt{http://astro.berkeley.edu/\~{}cooper/deep/spec2d/}} 
developed by the DEEP2 Redshift Survey Team \citep{davis2003}. 
We used a modified version of the \texttt{spec2d} (Capak et al. in prep.)
for the data taken with HDF11D, 
since those data were dithered. 
Wavelength calibration was achieved by fitting to the arc lamp emission lines.
The spectra were flux calibrated 
with the standard stars G191B2B and Wolf 1346. 
We applied no correction for slit-loss effects,
since the same slit width was used for both the standard stars and science targets. 
We estimate the uncertainty due to the slit-loss correction 
by comparing the fluxes of the two-dimensional spectra 
in the slit width with total fluxes 
and find this to be less than $10${\%}. 
We measure the $1 \sigma$ sky noises from the pixel distributions 
around $9600$ {\AA}, 
the wavelength corresponding to that of Ly$\alpha$ at $z \approx 6.9$, 
the expected peak of the redshift distribution 
for $z$-dropout selection \citep[Figure 6 of][]{ouchi2009b}.
In the measurements, 
we do not avoid the wavelength ranges 
significantly affected by strong OH lines.
We find the $1 \sigma$ sky noise to be ($1.4-5.7$) $\times 10^{-18}$ erg s$^{-1}$ cm$^{-2}$ 
(for details, see Section \ref{subsec:flux}).
We summarize the $1 \sigma$ flux limits in Table \ref{tab:sum_followup}.

\begin{deluxetable*}{cccccc} 
\tablecolumns{6} 
\tablewidth{0pt} 
\tablecaption{Summary of Follow-up Spectroscopy \label{tab:sum_followup}}
\tablehead{
    \colhead{Object}    & \colhead{Mask}  & \colhead{Total Exposure} & \colhead{flux limit ($1\sigma$)} & \colhead{$y^{({\rm total})}$} & \colhead{EW$_0^{{\rm Ly}\alpha}$ limit ($3\sigma$)}  \\
    \colhead{ } & \colhead{ } & \colhead{[sec]} & \colhead{[erg s$^{-1}$ cm$^{-2}$]} & \colhead{[mag]} & \colhead{[{\AA}]} 
}
\startdata 
GN-$152505$ 	& GNZD1B, HDF11C, HDF11D &  $39800$  & $3.0 \times 10^{-18}$ & $25.2$ & $10$ \\
GN-$108036$\tablenotemark{$\dagger$} 	& GNZD1B, HDF11C, HDF11D &  $39800$  & $4.2 \times 10^{-18}$ & $25.5$ & $7.1$ \\
SDF-$63544$\tablenotemark{$\dagger$} 			& SDFZD1B, SDFZD3 &  $49350$ & $2.1 \times 10^{-18}$ & $25.1$ & $5.8$ \\
SDF-$83878$ 			& SDFZD1B &  $19350$ & $3.3 \times 10^{-18}$ & $25.1$ & $12$ \\
SDF-$46975$\tablenotemark{$\dagger$} 			& SDFZD1B, SDFZD3 &  $49350$ & $1.9 \times 10^{-18}$ & $25.2$ & $7.1$   \\
SDF-$76507$ 			& SDFZD1B &  $19350$ & $2.4 \times 10^{-18}$ & $25.2$ & $9.2$ \\
SDF-$123919$ 		& SDFZD4 &  $7200$ & $5.3 \times 10^{-18}$ & $25.4$ & $31$  \\
SDF-$75298$ 			& SDFZD1B &  $19350$ & $1.4 \times 10^{-18}$ & $25.5$ & $6.5$ \\
SDF-$121418$ 		& SDFZD4 &  $7200$ & $4.5 \times 10^{-18}$ & $25.6$ & $31$ \\
SDF-$107344$ 		& SDFZD4 &  $7200$ & $5.7 \times 10^{-18}$ & $25.7$ & $53$  \\
SDF-$136726$ 		& SDFZD4 &  $7200$ & $5.1 \times 10^{-18}$ & $25.7$ & $45$ 
\enddata 
\tablecomments{
The flux limits of the objects with Ly$\alpha$ detections 
(GN-$108036$, SDF-$63544$, and SDF-$46975$) 
were estimated using the sky-noise distribution in the vicinity of the Ly$\alpha$ line, 
while those of objects lacking Ly$\alpha$ detections 
were estimated using the sky-noise distribution between $9400 - 9800$ {\AA}.   
The EW$_0^{{\rm Ly}\alpha}$ limits 
of the Ly$\alpha$-detected objects were estimated 
by dividing their Ly$\alpha$ flux limits 
by their UV continuum flux densities 
and $(1+z_{\rm spec})$, 
while those of objects without Ly$\alpha$ detection 
were estimated assuming their redshifts are equal to $6.9$.
}
\tablenotetext{$\dagger$}{%
$z$-dropout with Ly$\alpha$ detection.
}
\end{deluxetable*} 

\begin{deluxetable*}{cccccccccc} 
\tablecolumns{10} 
\tablewidth{0pt} 
\tablecaption{Properties of Spectroscopically Confirmed Ly$\alpha$-Emitting $z$-dropouts \label{tab:properties}}
\tablehead{
    \colhead{Object}    & \colhead{Redshift} 
    & \colhead{$f^{{\rm Ly}\alpha}$}  &  \colhead{$L^{{\rm Ly}\alpha}$}  & \colhead{$m_{\rm cont}$}  & \colhead{$M_{\rm cont}$}  & \colhead{EW$_0^{{\rm Ly}\alpha}$}    & \colhead{$S_w$} & \colhead{FWHM} & \colhead{$V_{\rm FWHM}$} \\
    \colhead{}    & \colhead{} 
    & \colhead{[erg s$^{-1}$ cm$^{-2}$]} & \colhead{[erg s$^{-1}$]} & \colhead{[mag]}    & \colhead{[mag]}    & \colhead{[{\AA}]}    & \colhead{[{\AA}]} & \colhead{[{\AA}]} & \colhead{[km s$^{-1}$]}
}
\startdata 
GN-108036
&  $7.213$
& $2.5 \times 10^{-17}$  &  $1.5 \times 10^{43}$ &  $25.2$ & $-21.8$  &  $33$  &  $4.1^{+0.7}_{-0.7}$ &  $15$  &  $442$  \\
SDF-63544
&  $6.965$
& $2.8 \times 10^{-17}$  &  $1.6 \times 10^{43}$  & $25.4$ & $-21.6$ & $43$  &  $12.6^{+0.4}_{-0.4}$  &  $6.7$  &  $207$  \\
SDF-46975
&  $6.844$
& $2.7 \times 10^{-17}$  &  $1.5 \times 10^{43}$ &  $25.4$  & $-21.5$ &  $43$  &  $8.6^{+0.6}_{-0.7}$ &  $12$  &  $374$ 
\enddata 
\end{deluxetable*} 

\begin{figure}
\begin{center}
   \includegraphics[scale=0.65]{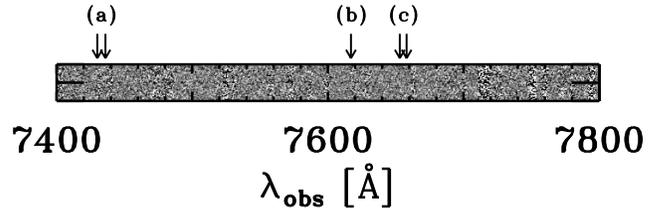}
\end{center}
 \caption[]
{
Two-dimensional spectrum of GN-108036 
in the wavelength range of $7400-7800${\AA}. 
The size along the spatial axis is $5.0${\arcsec}. 
No emission line is detected in this wavelength range, 
which argue against the possibility that 
the detected line at $\simeq 9980${\AA} is 
H$\alpha$ at $z = 0.521$. 
In this case, 
an [{\sc Oiii}]$\lambda 5007$ line would have been detected 
at approximately $7617${\AA} marked by an arrow (b).
No line detection also suggests that 
the detected line at $\simeq 9980${\AA} is not 
either an [{\sc Oiii}]$\lambda 5007$ at $z = 0.994$ 
or H$\beta$ at $z = 1.054$. 
In the former/latter case, 
the [{\sc Oii}]$\lambda\lambda 3726,3729$ doublet 
would have been detected at the position marked by arrows (a)/(c). 
}
\label{fig:blue_gn108036}
\end{figure}

\section{RESULTS} \label{sec:results}

\subsection{Redshift Identification} \label{subsec:redshift_id}

Figure \ref{fig:2d1dspectra} shows a prominent emission line 
in the spectra of SDF-63544 and SDF-46975. 
We fit Gaussian profiles to these lines using the IDL 
\verb|MPFIT| routine \citep{markwardt2009}\footnote{\texttt{http://www.physics.wisc.edu/\~{}craigm/idl/idl.html}},  
and find the emission lines correspond to 
the observed central wavelengths $9683${\AA} (SDF-63544) 
and $9536${\AA} (SDF-46975)\footnote{The best-fit central 
wavelengths are slightly shorter ($\sim 3$ {\AA}), 
if we fit an asymmetric Gaussian profile in which the standard 
deviation on the blue side is smaller than that on the red side, 
as a better approximation of the observed profiles 
but with an additional free parameter.}.
One $z$-dropout galaxy, SDF-63544, was previously identified 
as a Ly$\alpha$ emitter, IOK-1, whose 
redshift was confirmed spectroscopically by \cite{iye2006}.
However, the signal-to-noise ratio ($S/N$) of the Ly$\alpha$ line 
in their spectrum is only $\simeq 5$.  
We deemed it worthwhile to detect the line at higher $S/N$ 
and with higher resolution 
in order to verify the line detection
and identify the line based on line profile analysis. 
We confirm that 
the central wavelength of the emission line 
is almost the same as that derived previously ($9682${\AA}).

For one of the targets in GOODS-N, GN-108036, 
we detect an emission line 
at about $9980${\AA} (Figure \ref{fig:2d1dspectra_GN108036}). 
We confirm the line detection in three independent DEIMOS observations, 
performed in 2010 and 2011, with different configurations. 
In 2010, we obtained the spectrum 
using mask GNZD1B, 
the $830$ lines mm$^{-1}$ grating, 
and the OG550 filter 
(first two-dimensional spectrum in the top panel in Figure \ref{fig:2d1dspectra_GN108036}). 
In the 2011 run, 
we used a different set up with 
the $600$ lines mm$^{-1}$ grating, 
the GG455 filter,  
and a different mask (HDF11C) 
to locate the spectrum on a different position on the DEIMOS CCD 
(second one in the top panel in Figure \ref{fig:2d1dspectra_GN108036}). 
The HDF11D data were obtained over two nights of observing, 
and the line is detected in independent reductions of each night's data, 
as well as in the combined spectrum. 
We also used a third configuration for our 2011 run, 
with the $830$ lines mm$^{-1}$ grating, 
the OG550 filter, 
and an entirely new mask HDF11D 
(third one in the top panel in Figure \ref{fig:2d1dspectra_GN108036}). 
Although the grating and the filter here are 
the same as those used in the 2010 observation, 
the mask is different. 
The 2011 observations 
with the 830 grating were also dithered, 
whereas the 2010 observations with the 830 grating were undithered, 
and the pipeline processing was correspondingly different 
for the two data sets. 
The three observations of GN-108036 
independently confirm that 
the line is detected at three different positions 
on the DEIMOS CCDs, 
with two different spectroscopic setups. 
This strongly argues against the observed feature being an artifact. 
We fit Gaussian functions to the line profile in Figure \ref{fig:2d1dspectra_GN108036}, 
using the \verb|MPFIT| routine, 
and find a central wavelength of $9984${\AA}.

We investigate the possibility 
that the observed lines 
are something other than 
Ly$\alpha$ (i.e., H$\alpha$, H$\beta$,  [{\sc Oii}], or [{\sc Oiii}]). 
The main argument in favor of the observed lines being Ly$\alpha$ 
is their morphology and the clear asymmetry of the lines (see below). 
Furthermore, the lines are unlikely to be H$\beta$ or  [{\sc Oiii}] at $z \sim 0.9 - 1.0$, 
since, in this case, 
we expect to detect additional lines. 
If the detected line were H$\beta$, 
the [{\sc Oiii}]$\lambda 5007$ line would fall 
at $9972${\AA} for SDF-63544, 
$9822${\AA} for SDF-46975, and
$10284${\AA} for GN-108036. 
However, none of the objects show the corresponding detections,
implying the line ratio [{\sc Oiii}]$\lambda 5007$/H$\beta$ 
of $\lesssim 0.2$ -- $0.5$ ($3\sigma$ upper limit).
Galaxies with $12 + \log ({\rm O}/{\rm H}) \gtrsim 8.8$ -- 9.0 
meet these low ratios (e.g., Nagao et al. 2006), 
but our objects are unlikely to be such metal rich, 
because if they were, they should be very massive 
(from the mass-metallicity relation) and thus 
their broadband magnitudes would be much brighter than observed.
In the case that the detected line were [{\sc Oiii}]$\lambda 5007$, 
the H$\beta$ line would be seen 
at $9401${\AA} for SDF-63544, 
$9259${\AA} for SDF-46975, and
$9674${\AA} for GN-108036. 
However, none of the objects show the corresponding detections, 
implying the line ratio [{\sc Oiii}]$\lambda 5007$/H$\beta$ 
of $\gtrsim 2.0$ -- 4.0 ($3\sigma$ lower limit). 
These lower limits correspond to subsolar oxygen abundances
\citep[$\sim 7.0-8.7$; See Figure 17 of][]{nagao2006},
which may be typical of low-mass galaxies.
Thus, unfortunately the non-detection of H$\beta$ cannot 
strongly rule out the possibility of our objects 
being [{\sc Oiii}]$\lambda 5007$\ emitters.
In addition, the lines are unlikely to be [{\sc Oii}]. 
If the lines were [{\sc Oii}] emitters at $z \sim 1.6$, 
the FWHM resolution of $3.3${\AA} 
would have distinguished 
the two components of the doublet, 
separated by $\sim 7${\AA}.
In their Figure 10,  
\cite{hu2004} showed  
the spectra of emission-line objects with the [{\sc Oii}] 
doublet signature, which were 
obtained with Keck/DEIMOS 
using the same configuration as ours 
(the $830$ grating and the OG550 filter).  
The fact that 
the detected lines in our spectra are singlet 
strongly argues against
the possibility that they are $z \sim 1.6$ [{\sc Oii}] emitters. 
Moreover, the non-detection of the galaxies 
in the deep $i$-band images\footnote{The $2 \sigma$ limiting total magnitude 
of the SDF (GOODS-N) $i$-band image is $28.3$ ($27.6$).}
strongly disfavors the possibility that 
these galaxies are actually at $z \sim 1.6$. 
For GN-108036, 
if the detected line were H$\alpha$ at $z = 0.521$, 
then we might also expect to detect 
[{\sc Oiii}] or H$\beta$, 
which is actually not detected 
as shown in Figure \ref{fig:blue_gn108036}. 
For SDF-63544 and SDF-46975, 
we cannot check the detection of [{\sc Oiii}]$\lambda 5007$
since their spectra do not cover the wavelength range 
blueward of $7500${\AA}.

To quantify the asymmetry of the lines, 
we introduce the weighted skewness parameter, $S_w$, 
following \cite{kashikawa2006}. 
This can be used to distinguish Ly$\alpha$ from other emission lines. 
$S_w$ is defined as the product of the skewness 
(the third moment of flux distribution) and the width of the line. 
Ly$\alpha$ emission lines at high redshifts typically have large positive $S_w$ values, 
while other possible lines, H$\alpha$, H$\beta$,  and [{\sc Oiii}], 
are nearly symmetric and have almost zero values of $S_w$. 
When the [{\sc Oii}] line is not resolved, its $S_w$ value is expected to be small, 
since the [{\sc Oii}]$\lambda 3726$ line is weaker than [{\sc Oii}]$\lambda 3729$ \citep[e.g.,][]{rhoads2003}. 
The $S_w$ values of the three $z$-dropouts are 
estimated to be 
$S_w = 12.6 \pm 0.4${\AA} for SDF-63544,  
$S_w = 8.6^{+0.6}_{-0.7}${\AA} for SDF-46975, and 
$S_w = 4.1 \pm 0.7${\AA} for GN-108036, 
which means that all the three lines have an asymmetric profile 
with a sharp decline on the blue side and a long tail on the red side, 
as is commonly seen in Ly$\alpha$ at high redshifts \citep[e.g.,][]{shimasaku2006,kashikawa2006}.
\cite{kashikawa2006} empirically set $S_w = 3${\AA} as a critical value
to distinguish Ly$\alpha$ emission from other emission lines 
for galaxies at $z>5.7$
\citep[see also,][]{shimasaku2006}. 
Our objects have higher $S_w$ than the criterion, 
which would suggest that the detected lines are Ly$\alpha$.

Note that the asymmetric line profile in the spectra of 
GN-108036 might be caused by over-subtractions 
of OH lines blueward of the detected line. 
However, there is a window of OH lines near the detected line 
at the blue side. 
If the line had a symmetric profile, 
then we would expect to see faint emission in this window, 
similar to that seen on the red side of the line profile, 
but none is actually detected. 
The sky subtraction might affect the line profile to some extent, 
but the asymmetric line profile is likely to be real. 
In addition, 
there is no line detected other than the one at $\simeq 9980${\AA}. 
Furthermore, 
the galaxy is not detected 
in the deep optical broadband images 
from either Subaru or HST/ACS. 
Its colors 
($z-y > 1.6 \, (2 \sigma)$, $y - m_{\rm F140W} = 0.3$)
are consistent with those expected for 
a LBG at $z=7.2$. 
All of these factors 
support the interpretation that 
the detected line is redshifted Ly$\alpha$.

Thus, we conclude that all of the detected lines are redshifted Ly$\alpha$. 
The redshifts derived from the Ly$\alpha$ emission centroids are 
$z_{\rm spec} = 6.965$ for SDF-63544,  
$z_{\rm spec} = 6.844$ for SDF-46975, and 
$z_{\rm spec} = 7.213$ for GN-108036. 
These redshifts might be overestimated
since the Ly$\alpha$ emission lines of LBGs
are typically shifted redward of their systemic redshifts  
\citep[e.g.,][]{pettini2001,shapley2003,steidel2010}. 
The velocity offsets are typically less than $\sim 1000$ km s$^{-1}$, 
which corresponds to a difference between 
the systemic redshift ($z_{\rm sys}$) 
and that estimated from the Ly$\alpha$ line ($z_{{\rm Ly}\alpha}$) 
of 
$\Delta z = z_{\rm sys}  - z_{{\rm Ly}\alpha} \lesssim -0.027$.

{\sc Nv} $\lambda 1240$ is 
the only high-ionization metal line, 
indicative of AGN activity, 
which would fall within our spectral range. 
The line would fall at $9876${\AA} for SDF-63544, 
$9727${\AA} for SDF-46975, and
$10184${\AA} for GN-108036. 
None of the objects shows this emission line, 
placing a $3 \sigma$ lower limit to 
the line ratio Ly$\alpha$/{\sc Nv} of 
$\gtrsim 4$ for SDF-63544, 
$\gtrsim 5$ for SDF-46975, 
and $\gtrsim 2$ for GN-108036.
However, 
since a typical high-$z$ AGN has
a line ratio of Ly$\alpha$/{\sc Nv} $=4-20$ \citep[e.g.,][]{mccarthy1993}, 
these lower limits are mostly not useful to exclude the possibility 
that the objects are AGN hosts; to do so much deeper spectroscopy 
is needed.
Throughout this paper, we assume that the light of our objects 
is not contaminated by AGNs.

\subsection{Line Flux Measurement} \label{subsec:flux}

We compute the Ly$\alpha$ line flux 
by summing flux densities around the line, 
neglecting the minor contribution of UV continuum.
This corresponds to the range 
$9677-9699$ {\AA} for SDF-$63544$, 
$9526-9551$ {\AA} for SDF-$46975$, 
and $9977-9997$ {\AA} for GN-$108036$. 
The estimated fluxes are summarized in Table \ref{tab:properties}.

SDF-63544 was previously discovered by \cite{iye2006} 
who report Ly$\alpha$ fluxes of 
$2.0 \times 10^{-17}$ erg s$^{-1}$ cm$^{-2}$
and 
$2.7 \times 10^{-17}$ erg s$^{-1}$ cm$^{-2}$
from their spectrum and narrow-band image, respectively. 
Our estimated Ly$\alpha$ flux of SDF-63544 is  
$2.8 \times 10^{-17}$ erg s$^{-1}$ cm$^{-2}$,  
consistent with those previous measurements.

In order to estimate flux limits for the objects without Ly$\alpha$ detections, 
for each object we sample the one-dimensional spectra in $25${\AA} bins, 
comparable to the width of Ly$\alpha$ lines. 
We then estimate the $1 \sigma$ flux limit 
by fitting Gaussian functions to the flux distribution 
over the wavelength range $9400-9800${\AA}. 
The $1\sigma$ flux limits for individual objects 
are listed in Table \ref{tab:sum_followup}.

\subsection{Equivalent Width Measurement} \label{subsec:EW}

We do not directly detect continuum 
in the spectra of the $z$-dropout galaxies. 
We can only estimate their continuum flux densities 
from their $y$-band magnitudes 
measured over a $1.8$ arcsec diameter aperture
\citep[Table 3 of][]{ouchi2009b}, 
allowing for 
aperture corrections, 
IGM absorption 
and Ly$\alpha$ emission.

\begin{figure}
\begin{center}
   \includegraphics[scale=0.8]{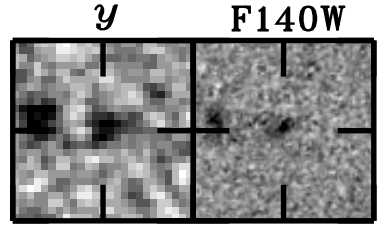}
\end{center}
 \caption[]
{
Images of GN-108036 
taken by the Subaru Suprime-Cam ($y$), 
and HST WFC3 (F140W). 
The size of each panel is $5${\arcsec}$\times 5${\arcsec}. 
North is up and east is to the left.
}
\label{fig:GN108036_faces}
\end{figure}

\begin{figure}
 \begin{center}
   \includegraphics[scale=0.7]{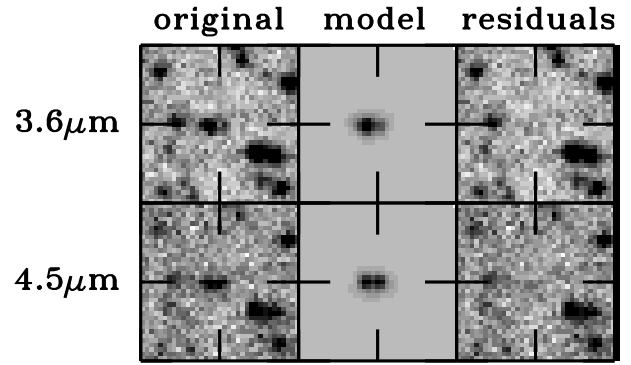}
\end{center}
 \caption[]
{
Images of GN-108036 in the IRAC $3.6\mu$m (top) and $4.5\mu$m (bottom). 
The panels from left to right 
show the original IRAC images, 
the best-fit model images constructed 
from PSF templates at the object positions 
measured from the HST WFC3 image, 
and the residuals after subtracting the model from the data.
The size of each panel is $20.4${\arcsec}$\times 20.4${\arcsec}. 
North is up and east is to the left.
}
\label{fig:irac_gn108036}
\end{figure}

\begin{figure}
   \includegraphics[scale=0.61]{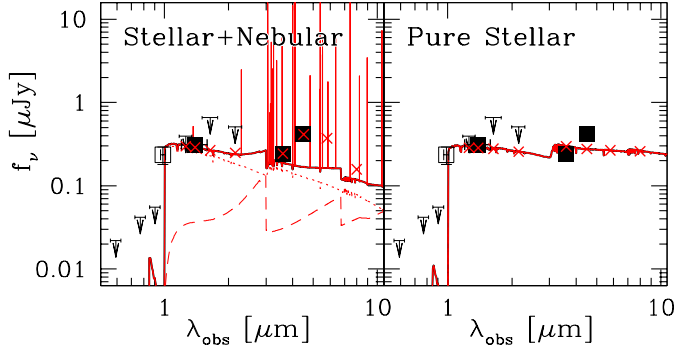}
 \caption[]
{
Results of SED fitting for the $z=7.213$ galaxy GN-108036. 
The filled squares show the observed flux densities 
used for SED fitting, 
while the open squares show those not used. 
The vertical arrows show $5 \sigma$ upper limits for 
HST/ACS $V_{606}$, $i_{775}$, $z_{850}$ \citep{ouchi2009b}, 
Subaru/MOIRCS $J$, $H$, and $K$ 
from left to right.
The left panel is 
for {\lq}stellar $+$ nebular{\rq} ($f_{\rm esc}^{\rm ion} = 0$) models. 
The red solid curve indicate the best-fit SED that is 
the sum of a stellar SED (red dotted curve) 
and a nebular SED (red dashed curve). 
The crosses indicate synthesized flux densities 
in individual bandpasses. 
The right panel is for {\lq}pure stellar{\rq} ($f_{\rm esc}^{\rm ion} = 1$) models. 
}
\label{fig:stellarpops}
\end{figure}

\begin{deluxetable*}{cccccccc} 
\tablecolumns{8} 
\tablewidth{0pt} 
\tablecaption{SED Fitting Results for $z=7.213$ galaxy\label{tab:stellarpops}}
\tablehead{
    \colhead{model}    & \colhead{$Z$} & \colhead{$\log M_{\rm star}$}    & \colhead{$E(B-V)_\star$} & \colhead{$\log$(Age)}    
    & \colhead{$\log$(SFR)}    & \colhead{$\log$(SSFR)}  & \colhead{$\chi^2$} \\
    \colhead{}    & \colhead{[$Z_\odot$]}    & \colhead{$[M_\odot]$}    & \colhead{[mag]} & \colhead{[yr]}    
    & \colhead{[$M_\odot$ yr$^{-1}$]}  & \colhead{[yr$^{-1}$]}  & \colhead{}
}
\startdata 
stellar $+$ nebular
& $0.2$  & $8.76^{+0.10}_{-0.12}$ &  $0.05^{+0.02}_{-0.03}$  
&  $6.76^{+0.10}_{-0.12}$  &  $2.00^{+0.02}_{-0.01}$  &  $-6.76^{+0.14}_{-0.17}$  &  $12.8$ 
\\
pure stellar
& $0.2$  & $9.35^{+0.12}_{-0.11}$ &  $0.11^{+0.01}_{-0.01}$
&  $7.38^{+0.13}_{-0.12}$  &  $2.00^{+0.02}_{-0.02}$  &  $-7.35^{+0.23}_{-0.20}$  &  $23.8$ 
\enddata 
\end{deluxetable*} 

To estimate the aperture correction for our $z$-dropouts, 
we select bright, non-saturated and isolated sources in the $y$-band images, 
and perform multi-aperture photometry 
to construct the curve of growth. 
We then determine the aperture correction 
by measuring the difference 
between magnitude over the $1.8$ arcsec diameter aperture 
and the total magnitude (the asymptote of the growth curve). 
The median aperture correction is $0.344$ mag for GOODS-N 
and $0.349$ mag for the SDF.

The flux in the $y$-band 
is affected by Ly$\alpha$ emission 
and IGM absorption shortward of rest-frame $1216${\AA}. 
We correct for the contribution due to Ly$\alpha$ emission 
using the estimated Ly$\alpha$ flux from section \ref{subsec:flux} 
for objects with strong Ly$\alpha$ emission, 
and the $3\sigma$ flux limit 
for the $z$-dropouts without Ly$\alpha$ detection. 
The IGM absorption is taken into account 
on the assumption of flat UV continuum in $f_\nu$
and 
using \cite{madau1995}. 
The $3 \sigma$ EW limits for the $11$ galaxies in the spectroscopic sample 
are listed in Table \ref{tab:sum_followup}. 
In Table \ref{tab:properties}, 
we list our estimates for the continuum magnitude 
$m_{\rm cont}$ at the Ly$\alpha$ wavelength, 
and the corresponding EWs 
that we derived for the Ly$\alpha$ emission lines.

\subsection{Photometry and Stellar Population Properties of the $z=7.213$ galaxy} \label{subsec:stellarpops}

The Subaru $y$-band photometry for the $z = 7.2$ galaxy GN-108036 
is difficult to interpret
because the flux measurement is strongly affected 
both by the Ly$\alpha$ emission line 
and
the Ly$\alpha$ forest IGM absorption, 
which should suppresses roughly $65${\%} 
of the intrinsic continuum flux from the galaxy 
within the $y$ bandpass.   
Fortunately, GN-108036 has been imaged 
with the HST WFC3-IR camera 
(Weiner et al.\ in preparation; HST program {\#}11600).
The image has a total exposure time of $1217$~sec 
and was obtained with the F140W filter, 
whose central wavelength is approximately $1.4\mu$m, 
sampling $1700${\AA} in the rest-frame UV continuum.
The Suprime-Cam $y$-band 
and WFC3 F140W images of GN-108036 are shown in Figure \ref{fig:stellarpops}. 
Photometry in the HST image 
measures a total magnitude of $25.17 \pm 0.07$~mag.
GN-108036 was also observed 
in the Subaru MOIRCS $JHK$ survey 
of the GOODS-N field \citep{kajisawa2011}. 
The galaxy is not significantly detect 
in any of these MOIRCS bands, 
with $5\sigma$ photometric upper limits of 
$24.9$ ($J$), $24.4$ ($H$), and $24.6$ ($K$),
marginally consistent with the WFC3 photometry.

The WFC3 F140W magnitude of GN-108036, 
$m = 25.17$, translates to a rest-frame $1700${\AA} luminosity 
$M_{1700} = -21.81$ at $z = 7.213$.  
There have been many recent estimates 
of the UV luminosity function at $z = 7$ 
\citep[e.g.,][]{ouchi2009b,oesch2010,mclure2009,grazian2011}.
These have consistently found values 
of the characteristic luminosity $M^\ast$ 
in the range $-19.9$ to $-20.3$.  
GN-108036 is therefore an impressively luminous galaxy, 
with $L_{\rm UV} \approx 4$ to 6$L^\ast$;  
indeed, it is roughly twice as bright as 
an $L^\ast_{\rm UV}$ LBG at $z = 2$ to $3$ 
\citep[$M^\ast_{\rm UV} \approx -21.0$,][]{reddy2008}.
The exponential cut-off of a Schechter luminosity function 
would imply that galaxies this luminous would be rare indeed. 
For example, the best-fit Schechter parameterization 
of the $z = 7$ luminosity function 
from \cite{bouwens2011} 
yields a space density 
$2.7 \times 10^{-7}$~Mpc$^{-3}$ 
for galaxies with $M_{\rm UV} \leq -21.81$. 
For a redshift interval $\Delta z = 1$ at $z = 7$, 
we would expect to find only $0.1$ galaxies 
this luminous within the $160$~arcmin$^2$ 
GOODS-N ACS/WFC3/IRAC field, 
or about $1$ galaxy over the $1568$~arcmin$^2$ 
covered by our Suprime-Cam survey 
in the combined SDF and GOODS-N fields. 
In fact, the redshift selection function for $z$-dropouts
in our $z-y$ survey 
is significantly narrower than $\Delta z = 1$ 
\citep[see][]{ouchi2009b}, 
making the large UV luminosity of GN-108036 
still more remarkable.   
It is impossible to judge based on {\em a posteriori} 
statistics from a single object, 
but either we were quite lucky to find 
and spectroscopically confirm such a bright galaxy, 
or perhaps 
the bright end of the UV luminosity function 
at $z = 7$ has
been underestimated in studies to date.  
In fact, the observational constraints 
on the bright end of the luminosity function
come from just a few ground-based surveys, 
including our own \citep{ouchi2009b} and
that of \cite{castellano2010,castellano2010b}. 
Evidently more deep imaging and spectroscopic studies 
over wider fields are needed 
to provide better measurements. 
The $y$-band itself becomes an unreliable indicator 
of luminosity for galaxies at $z > 7$ 
without exact measurements of the galaxy redshift 
and Ly$\alpha$ line flux, 
so deep near-infrared data at longer wavelengths 
(such as that from HST WFC3) will be essential 
for a robust determination of the luminosity function.

GN-108036 also falls within the field 
covered by the extremely deep IRAC imaging 
from the GOODS Spitzer Legacy program (PI: M. Dickinson). 
As was noted in \cite{ouchi2009b}, 
the galaxy is faintly detected 
in the IRAC $3.6\mu$m and $4.5\mu$m images, 
which roughly sample the rest-frame $B$ and $V$ band light at $z = 7.2$, 
although it is partially blended with a foreground galaxy 
located $1${\arcsec} 
to the east (Figure \ref{fig:irac_gn108036}).  
In order to extract reliable IRAC fluxes, 
we measure the positions of the two galaxies 
in the WFC3 F140W image, 
and then position unit-normalized IRAC PSF images 
at these locations in the background-subtracted
$3.6\mu$m and $4.5\mu$m images.  
We then fit these PSF templates to the IRAC data, 
minimizing $\chi^2$ to derive the best-fitting fluxes 
and their uncertainties.  
The middle panels in Figure \ref{fig:irac_gn108036} 
show the best-fit model 
and the right panels are the residual images.  
In this way, 
we obtain the total magnitudes of GN-108036 of 
$m_{3.6\mu{\rm m}} = 25.44 \pm 0.13$ 
and $m_{4.5\mu{\rm m}} = 24.86 \pm 0.13$,
respectively.

We fit models to the photometry of GN-108036 
to infer its stellar population properties. 
The procedure is the same as that of \cite{ono2010}, 
except that redshift is fixed at $z = 7.213$.  
We use the stellar population synthesis model 
GALAXEV \citep{bc2003} 
for the stellar component of the SEDs, 
and
consider two extreme cases for nebular emission: 
$f^{\rm ion}_{\rm esc} = 0$, 
where ionizing photons are all converted 
into nebular emission (the {\lq}stellar $+$ nebular{\rq} case),
and 
$f^{\rm ion}_{\rm esc} = 1$, 
where all ionizing photons escape 
from the galaxy (the {\lq}pure stellar{\rq} case).  
Nebular spectra (lines and continua) 
are calculated 
following the procedure given in \cite{schaerer2009}.  
We adopt the Salpeter initial mass function \citep[IMF;][]{salpeter1955},  
a constant rate of star formation, 
and 
stellar and gas metallicities $Z = 0.2 Z_\odot$.  
For dust attenuation, we use the functional form of \cite{calzetti2000} 
with the assumption that 
$E(B-V)_{\rm gas} = E(B-V)_\star$, 
as proposed by \cite{erb2006c}. 
The model SEDs are fit to the observed flux densities 
in the WFC3 F140W and
IRAC 3.6$\mu$m and 4.5$\mu$m bands, 
and 
its MOIRCS $JHK$ magnitudes.
We do not use the $y$-band photometry 
since it is strongly affected by IGM attenuation 
and Ly$\alpha$ emission.  
The free parameters in the fitting are 
stellar mass, age,
and dust extinction, 
with three degrees of freedom.

The results of SED fitting are summarized in Table \ref{tab:stellarpops}, 
and the best-fitting SEDs are shown in Figure \ref{fig:stellarpops}.  
The best-fit models have small stellar masses, 
$4 \times 10^8 M_\odot$ to $3 \times 10^9 M_\odot$, 
very young ages, $4$ to $32$ Myr, 
and modest color excesses $E(B-V)_\star = 0.02$ to $0.12$\footnote{The uncertainties 
should be larger if star-formation history and metallicity are varied, 
although we fix them due to the small number of the data points.}.
Deeper $K$-band photometry would be needed 
to reliably constrain the amplitude of the Balmer break, 
but the blue rest-frame UV-to-optical color 
$m_{\rm F140W} - m_{3.6\mu{\rm m}} = -0.27$ 
suggests that the break is small 
and that the light from the galaxy is 
dominated by very young and relatively unreddened stars.

If the flux in the IRAC bands 
has a significant contribution from nebular emission, 
then the intrinsic stellar SED would be even bluer 
than the directly measured WFC3-to-IRAC color
would suggest (see Figure \ref{fig:stellarpops}, left). 
There is evidence that this is the case.  
The very red IRAC $3.6\mu$m to $4.5\mu$m color 
is not easily reproduced by stellar emission alone, 
but can be explained by a strong contribution 
of nebular emission in the $4.5\mu$m bandpass.  
At $z = 7.2$, 
the IRAC $3.6\mu$m band is mostly free of 
strong emission lines, 
but the IRAC $4.5\mu$m band includes 
[{\sc Oiii}]$\lambda 4959$  
and $\lambda 5007$ (H$\beta$ is largely excluded).  
Several recent studies have suggested that 
strong nebular emission lines can significantly 
affect IRAC photometry for high redshift galaxies 
\citep[e.g.,][]{schaerer2009,vanzella2010,raiter2010,ono2010,shim2011}, 
and galaxies with extremely strong [{\sc Oiii}] 
(${\rm EW}_0 > 1000$\AA) have been 
identified at $z < 1$ \citep[e.g.,][]{kakazu2007}
and $z = 1$ to 1.5 \citep{atek2010}.
Here we see evidence for strong [{\sc Oiii}] at $z = 7.2$;  
in the best-fit stellar$+$nebular model, 
the [{\sc Oiii}] lines contribute $\sim 60${\%} 
of the flux in the IRAC $4.5\mu$m band.

Using the conversion 
from UV continuum luminosity density to star formation rate
from \cite{madau1998} for a Salpeter IMF, 
and assuming no extinction,
the $1700${\AA} luminosity of GN-108036 
implies a star formation rate (SFR) of $29 M_\odot$ yr$^{-1}$. 
However, the best-fitting stellar population models 
give larger SFR $ \approx 100 M_\odot$ yr$^{-1}$ 
(Table \ref{tab:stellarpops}).  
In part this is due to extinction in the models, 
and in part to the very young ages 
that are implied by the SED-fitting.   
The standard conversion factors of UV luminosity 
per unit SFR \citep[e.g., from][or \citealt{kennicutt1998}]{madau1998}
assume star formation timescales $\gtrsim 10^8$ years. 
At younger ages, the UV continuum is still building 
toward its steady-state value, 
and SFRs derived using the standard conversion factors 
will be underestimated.  
For comparison, the Ly$\alpha$ line luminosity 
($1.5 \times 10^{43}$ erg s$^{-1}$) 
yields another estimate of SFR(Ly$\alpha) = 13.6 M_\odot$ yr$^{-1}$, 
again adopting a standard Salpeter IMF conversion factor \citep{kennicutt1998} 
and assuming Case B recombination
for the H$\alpha$/Ly$\alpha$ flux ratio.  
This SFR is significantly smaller than either 
estimate from the continuum SED, 
as expected since Ly$\alpha$ is subject to 
strong attenuation from dust extinction and by the IGM.

Because the unusual IRAC color of GN-108036 
strongly favors the results 
from the {\lq}stellar$+$nebular{\rq} model fitting, 
we conclude that the stellar mass of this galaxy is 
most likely smaller than $10^9 M_\odot$.  
This value is not atypical compared to 
other published estimates for galaxies at $z \sim 7$ 
\citep[e.g.,][]{finkelstein2009f,gonzalez2011}. 
However, when combined with the bright UV luminosity 
and 
the large derived star formation rate, 
it implies an extremely high specific star formation rate 
(SSFR) $> 10^{-7}$ yr$^{-1}$.   
This is at least $50$ times higher than 
the mean values that have been estimated 
for samples of LBGs at $4 < z < 7$ \citep{gonzalez2009}, 
and suggests that GN-108036 is seen 
at a special moment in its life cycle 
when it is undergoing a very strong starburst, 
presumably with a short duration.

\begin{figure*}
\hspace{0.3cm}
   \includegraphics[scale=0.6]{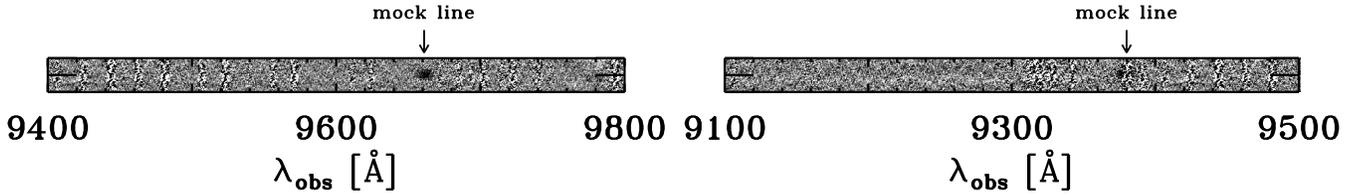}
 \caption[]
{
Examples of mock spectra 
made by artificially adding a mock emission line 
to the two-dimensional spectrum of GN-152505. 
The mock emission line 
has Ly$\alpha$ equivalent width of $50${\AA} and continuum magnitude of $25.57$. 
The left panel shows a spectrum 
whose mock line is found at $\simeq 9660${\AA}, 
while a mock line ($\simeq 9380${\AA}) is not recovered in the right spectrum 
due to the residuals of severe OH lines around the artificial line. 
}
\label{fig:examples_simuspec}
\end{figure*}

Although the IRAC color favors 
{\lq}stellar$+$nebular{\rq} modeling, 
the very young age derived may be in conflict with 
a simple expectation that the duration of star formation 
cannot be shorter than the dynamical time of the system, 
since it will take around the dynamical time 
to have the gas of the system collapse to form stars.
GN-108036 has an FWHM of $0.30''$ in the F140W image and 
a Ly$\alpha$ line width of $4.4 \times 10^2$ km s$^{-1}$. 
Dividing the former by the latter gives a rough estimate 
of the dynamical time of $\sim 4$ Myr, 
which is marginally consistent with the age from SED fitting. 
Although this possible problem is not directly related to 
the discussion and conclusions of this paper, 
similar conflicts between stellar age and dynamical time 
are found in other papers
\citep[e.g.,][]{schaerer2009,schaerer2010,ono2010}, 
highlighting the need for further research.

Assuming a flat $f_{\nu}$ UV spectral slope, 
we derive an equivalent width for Ly$\alpha$ 
EW$_0^{{\rm Ly}\alpha} = 33${\AA}.   
Another galaxy from our sample, SDF-63544, 
was also observed with HST/WFC3 
in the F125W, F140N, and F160W 
bandpasses \citep[][Egami et al. in preparation]{cai2011}.  
A fit to this photometry 
yields a UV continuum slope 
$f_{\lambda} \propto \lambda^{-2.46}$ \citep{cai2011}, 
and
a continuum flux magnitude of $25.5 \pm 0.1$ at $9683${\AA}, 
the wavelength of Ly$\alpha$, 
consistent with our estimate 
from the $y$-band photometry in Section \ref{subsec:EW}.

\begin{deluxetable*}{ccccccc} 
\tablecolumns{7} 
\tablewidth{0pt} 
\tablecaption{Properties of $z$-dropouts Spectroscopically Observed in the Previous Studies \label{tab:previous}}
\tablehead{
    \colhead{Object}    & \colhead{Redshift\tablenotemark{a}} & \colhead{$m_{\rm cont}$\tablenotemark{b}} & \colhead{$M_{\rm UV}$\tablenotemark{c}} 
    & \colhead{$f^{{\rm Ly}\alpha}$~\tablenotemark{d}}  & \colhead{EW$_0^{{\rm Ly}\alpha}$} & \colhead{comments} \\
    \colhead{}    & \colhead{} & \colhead{[mag]} & \colhead{[mag]}
    & \colhead{[erg s$^{-1}$ cm$^{-2}$]} & \colhead{[{\AA}]}  & \colhead{} 
}
\startdata 
\multicolumn{7}{c}{\cite{fontana2010}} \\ \hline 
G2\_1408 & $6.972$ & $26.37$ & $-20.49$ &  $3.4 \times 10^{-18}$ &  $13$ &  $S/N({\rm Ly}\alpha)=7$, \cite{castellano2010}, $Y_{\rm OPEN}$ Hawk-I \\
G2\_2370 & $6.8$ & $25.56$ & $-21.27$ & $<2.5 \times 10^{-18}$ &  $<4.8$ &  \cite{castellano2010}, $Y_{\rm OPEN}$ Hawk-I \\
G2\_4034 & $6.8$ & $26.35$ & $-20.50$ & $<2.5 \times 10^{-18}$ &  $<9.7$ &  \cite{castellano2010}, $Y_{\rm OPEN}$ Hawk-I \\
G2\_6173 & $6.8$ & $26.53$ & $-20.33$ & $<2.5 \times 10^{-18}$ &  $<11$ & \cite{castellano2010}, $Y_{\rm OPEN}$ Hawk-I  \\
H\_9136 & $6.8$ & $25.90$ & $-20.94$ & $<2.5 \times 10^{-18}$ &  $<6.4$ &  \cite{hickey2010}, $Y_{\rm OPEN}$ Hawk-I \\
W\_6 & $6.8$ & $26.93$ & $-20.38$ & $<2.5 \times 10^{-18}$ &  $<11$ &  \cite{wilkins2010}, $Y_{098}$ WFC3-ERS \\
O\_5 & $6.8$ & $27.52$ & $-19.67$ & $<2.5 \times 10^{-18}$ &  $<21$ & \cite{oesch2010}, $Y_{105}$ WFC3-HUDF \\
\hline 
\multicolumn{7}{c}{\cite{vanzella2010d}} \\ \hline 
BDF-521 & $7.008$ & $25.86$ & $-20.63$ &  $1.62 \times 10^{-17}$ &  $64$ &  $S/N({\rm Ly}\alpha)=18$   \\
BDF-3299 & $7.109$ & $26.15$ & $-20.56$ & $1.21 \times 10^{-17}$ &  $50$ &  $S/N({\rm Ly}\alpha)=16$  \\
\hline 
\multicolumn{7}{c}{\cite{schenker2011}} \\ \hline 
ERS5847 & $6.48$\tablenotemark{$\dagger$} & $26.6$ & $-20.22$ & --- & --- & --- \\
ERS7376 & $6.79$\tablenotemark{$\dagger$} & $27.0$ & $-19.89$ & --- & --- & --- \\
ERS7412 & $6.38$\tablenotemark{$\dagger$} & $27.0$ & $-19.79$ & --- & --- & --- \\
ERS8119 & $6.78$\tablenotemark{$\dagger$} & $27.1$ & $-19.79$ & --- & --- & --- \\
ERS8290 & $6.52$\tablenotemark{$\dagger$} & $27.1$ & $-19.73$ & --- & --- & with a close neighbor \\
ERS8496 & $6.441$ & $27.3$ & $-19.51$ & $9.1 \times 10^{-18}$ & $65$ & $S/N({\rm Ly}\alpha)>5$ \\
ERS10270 & $7.02$\tablenotemark{$\dagger$} & $27.4$ & $-19.54$ & --- & --- & --- \\
ERS10373 & $6.44$\tablenotemark{$\dagger$} & $27.4$ & $-19.41$ & --- & --- & --- \\
A1703\_zD1 & $6.75$\tablenotemark{$\dagger$} & $24.1$ & $-20.39$ & --- & --- & Magnification factor $\mu = 9.0$\tablenotemark{$\ddagger$} \\
A1703\_zD3 & $6.89$\tablenotemark{$\dagger$} & $25.5$ & $-19.25$ & --- & --- & Magnification factor $\mu = 7.3$\tablenotemark{$\ddagger$} \\
A1703\_zD6 & $7.045$ & $25.8$ & $-19.36$ & $2.8 \times 10^{-17}$ & $65$ & Magnification factor $\mu = 5.2$\tablenotemark{$\ddagger$}, $S/N({\rm Ly}\alpha)>5$ \\
A1703\_zD7 & $8.80$\tablenotemark{$\dagger$} & $26.8$ & $-18.74$ & --- & --- & Magnification factor $\mu = 5.0$\tablenotemark{$\ddagger$} \\
A2261\_1 & $7.81$\tablenotemark{$\dagger$} & $26.9$ & $-18.85$ & --- & --- & Magnification factor $\mu = 3.5$ \\
BoRG\_4 & $8.27$\tablenotemark{$\dagger$} & $25.8$ & $-21.39$ & --- & --- & --- \\
EGS\_K1 & $8.27$\tablenotemark{$\dagger$} & $25.3$ & $-21.89$ & --- & --- & --- \\
HUDF09\_799 & $6.88$\tablenotemark{$\dagger$} & $27.7$ & $-19.21$ & --- & --- & --- \\
HUDF09\_1584 & $7.17$\tablenotemark{$\dagger$} & $26.7$ & $-20.27$ & --- & --- & --- \\
HUDF09\_1596 & $6.905$ & $26.8$ & $-20.12$ & --- & $30$ & --- \\
MS0451-03\_10 & $7.50$\tablenotemark{$\dagger$} & $26.7$ & $-16.10$ & --- & --- & Magnification factor $\mu = 50$ \\
\hline 
\multicolumn{7}{c}{\cite{pentericci2011}} \\ \hline 
NTTDF-474 & $6.623$ & $26.50$ & $-20.35$ & $3.2 \times 10^{-18}$ & $16$ & $S/N({\rm Ly}\alpha)=7$ \\
NTTDF-1479 & $6.8$ & $26.12$ & $-20.77$ & --- & ---\tablenotemark{$\ast$} & --- \\
NTTDF-1632 & $6.8$ & $26.44$ & $-20.45$ & --- & ---\tablenotemark{$\ast$} & --- \\
NTTDF-1917 & \textbf{---} & $26.32$ & --- & --- & ---\tablenotemark{$\ast$} & likely an interloper \\
NTTDF-2916 & $6.8$ & $26.64$ & $-20.25$ & --- & ---\tablenotemark{$\ast$} & --- \\
NTTDF-6345 & $6.701$ & $25.46$ & $-21.41$ & $7.2 \times 10^{-18}$ & $15$ & $S/N({\rm Ly}\alpha)=11$ \\
NTTDF-6543 & $6.8$ & $25.75$ & $-21.14$ & --- & ---\tablenotemark{$\ast$} & --- \\
BDF4-2687 & $6.8$ & $26.15$ & $-20.74$ & --- & ---\tablenotemark{$\ast$} & --- \\
BDF4-2883 & $6.8$ & $26.15$ & $-20.74$ & --- & ---\tablenotemark{$\ast$} & --- \\
BDF4-5583 & $6.8$ & $26.65$ & $-20.24$ & --- & ---\tablenotemark{$\ast$} & --- \\
BDF4-5665 & $6.8$ & $26.64$ & $-20.25$ & --- & ---\tablenotemark{$\ast$} & --- 
\enddata 
\tablecomments{
The upper limits of Ly$\alpha$ flux and equivalent width are $5\sigma$. 
}
\tablenotetext{a}{%
The redshift of objects without spec-$z$ is set at $6.8$. 
}
\tablenotetext{b}{%
For the objects of \cite{fontana2010}, \cite{vanzella2010d}, and \cite{pentericci2011}, 
$Y$-band magnitudes are taken from their Table 1, respectively.
For the objects of \cite{schenker2011}, 
$J_{125}$ magnitudes are taken from their Table 1.
}
\tablenotetext{c}{%
Objects from \cite{fontana2010} are taken 
from their Table 1. 
For the objects of \cite{vanzella2010d}, 
it is calculated from the $Y$-band magnitude corrected for 
Ly$\alpha$ contribution and IGM attenuation, $Y_{\rm cont}$. 
For the objects of \cite{schenker2011} and \cite{pentericci2011}, 
it is calculated from the continuum magnitude $m_{\rm cont}$ and redshift. 
The magnification factors are considered for the \cite{schenker2011} objects.
}
\tablenotetext{d}{%
For the objects of \cite{fontana2010}, 
the upper limit is estimated from the $1\sigma$ limiting flux 
at $9485${\AA}, corresponding to $z \sim 6.8$ Ly$\alpha$, 
around which the redshift distribution of their $z$-dropouts peaks 
\citep[Figure 1 in][]{castellano2010}. 
}
\tablenotetext{$\dagger$}{%
The photometric redshift estimated by 
\cite{schenker2011} or 
\cite{mclure2011}.
}
\tablenotetext{$\ddagger$}{%
These are estimated by \cite{bradley2011}.
}
\tablenotetext{$\ast$}{%
The EW detection limits are well below EW$=25${\AA} \citep{pentericci2011}.
}
\end{deluxetable*} 

\begin{deluxetable}{ccc} 
\tablecolumns{3} 
\tablewidth{0pt} 
\tablecaption{Summary of the samples \label{tab:summary_numbers}}
\tablehead{
    \colhead{ }    & \colhead{EW$_0^{{\rm Ly}\alpha} > 25${\AA}} & \colhead{EW$_0^{{\rm Ly}\alpha} > 55${\AA}}
}
\startdata 
\multicolumn{3}{c}{$-21.75 < M_{\rm UV} < -20.25$} \\ \hline 
This Study & $2/6$\tablenotemark{$\dagger$} & $0/10$ \\
\cite{fontana2010} & $0/6$ & $0/6$ \\
\cite{vanzella2010d} & $2/2$ & $1/2$ \\
\cite{schenker2011} & $0/2$ & $0/2$ \\
\cite{pentericci2011} & $0/7$ & $0/7$ \\
\hline 
\multicolumn{3}{c}{$-20.25 < M_{\rm UV} < -18.75$} \\ \hline 
\cite{fontana2010} & $0/1$ & $0/1$ \\
\cite{schenker2011} & $3/12$ & $2/12$ \\
\cite{pentericci2011} & $0/3$ & $0/3$ 
\enddata 
\tablecomments{
For the sample of \cite{schenker2011}, 
objects at $6.3 < z < 7.3$ are considered.
}
\tablenotetext{$\dagger$}{%
In our sample, 
four objects have Ly$\alpha$ EW limits larger than $25${\AA}. 
}
\end{deluxetable} 

\subsection{Ly$\alpha$ Fraction} \label{subsec:X_Lya}

In order to infer the ionizing state of the IGM, 
we study the evolution of the Ly$\alpha$ fraction 
by measuring its value at $z \sim 7$ 
and comparing it with previous measurements at $4<z<6$. 
For $4<z<6$ dropout galaxies, 
\cite{stark2010b} 
divided their sample into two UV luminosity bins, 
$-21.75 < M_{\rm UV} < -20.25$ 
and 
$-20.25 < M_{\rm UV} < -18.75$, 
and 
estimated their Ly$\alpha$ fractions. 
Since the UV absolute magnitudes of our $z$-dropouts 
are in the brighter range, 
in the following analysis for our sample 
we focus only on the Ly$\alpha$ fraction 
of dropout galaxies at $-21.75 < M_{\rm UV} < -20.25$.

We define the Ly$\alpha$ fraction 
as the ratio of the number of $z$-dropout galaxies with strong Ly$\alpha$ emission 
measured from their spectra 
to the total number of spectroscopically observed $z$-dropouts, 
\begin{equation}
X_{{\rm Ly}\alpha}^{{\rm EW}_c} 
	= \sum_i \alpha_i^{{\rm Ly}\alpha} p_i / \sum_i p_i, 
\end{equation}
where 
$\alpha_i^{{\rm Ly}\alpha}$ is $1$ 
if any $i$-th galaxy has a Ly$\alpha$ EW larger than 
a critical EW (EW$_c$) and $0$ otherwise. 
$p_i$
is the probability that the Ly$\alpha$ wavelength of the $i$-th 
$z$-dropout candidate is within the observable wavelength range - 
i.e., the range not contaminated by OH airglow lines, 
and is given by 
$p_i = \int C_i'(z) dz / \int C(z) dz$, 
where 
$C(z)$ 
is the probability distribution for $z$-dropouts 
as a function of redshift 
and 
$C_i'(z)$ 
is the effective redshift probability distribution function of 
$z$-dropout candidates when corrected for OH emission. 
Such a statistical analysis of the $p_i$ parameter is needed 
since, 
at the wavelength range $\sim 9000 - 10100$ {\AA}, 
where we expect to detect their Ly$\alpha$ line, 
our $z \sim 7$ candidates may not always satisfy the EW$_c$ criterion 
(i.e., $p_i = 1$), 
as the atmospheric OH airglow lines 
will significantly affect the spectrum in that wavelength range.

In order to compute $C'(z)$, 
we perform Monte Carlo simulations 
by generating mock Ly$\alpha$ emission lines 
with EW$_0 = 50${\AA} or $100${\AA} 
and $y$-band magnitudes of $25.24$ and $25.57$ mag.
These magnitudes correspond to 
the central values of two bins 
into which the $y$-band total magnitudes of our $z$-dropouts are divided 
in order of their brightness. 
We add the mock Ly$\alpha$ lines 
to an observed two-dimensional spectrum, 
following the redshift distribution, 
$C(z)$ \citep[Figure 6 in][]{ouchi2009b}. 
We then inspect the mock two-dimensional spectrum 
searching for a Ly$\alpha$ feature. 
Figure \ref{fig:examples_simuspec} shows examples of 
two-dimensional spectra apparently with and without
Ly$\alpha$ detected (left and right plots, respectively).
We perform a number of trials 
and evaluate the recovery rate of mock Ly$\alpha$ lines. 
The $C'(z)$ is then estimated by computing the weighted mean of Ly$\alpha$ EWs, 
assuming this to be a Gaussian function \citep{ouchi2008}.
We find that over $90${\%} of the mock Ly$\alpha$ lines are 
successfully recovered 
in our extremely deep spectra (i.e., $C_i'(z) \simeq C(z)$). 
Therefore, 
we choose not to correct for this effect in the following analysis. 
When computing the Ly$\alpha$ emission fraction, $X_{{\rm Ly}\alpha}^{\rm{EW}_c}$ 
above a certain equivalent width threshold EW$_c$, 
we divide the number of galaxies with {\em detected} EW(Ly$\alpha) > $EW$_c$ 
by the number of galaxies for which the 3$\sigma$ upper limits 
to the EW detectability 
are smaller than EW$_c$ 
(i.e., those galaxies for which Ly$\alpha$ lines 
stronger than EW$_c$ could have been detected).

From the sample of $11$ $z$-dropout galaxies with spectroscopic data studied here, 
we detect Ly$\alpha$ emission lines for $3$ sources, 
all with Ly$\alpha$ EWs larger than $25${\AA}. 
However, we do not include GN-108036 at $z=7.2$, 
since it is brighter than $-21.75$ in the rest-frame UV. 
We find $4$ objects with Ly$\alpha$ EW limits larger than this, 
giving a Ly$\alpha$ fraction, 
$X_{{\rm Ly}\alpha}^{25} = 33 \pm 27${\%} ($2/6$). 
If we set the 
EW$_c = 55${\AA}, 
the $2$ objects with Ly$\alpha$ detection 
have smaller EW$_0^{{\rm Ly}\alpha}$ than the criterion. 
The Ly$\alpha$ EW upper limits of $8$ $z$-dropout candidates 
without Ly$\alpha$ detection are lower than the EW$_c$. 
We thus obtain an upper limit of Ly$\alpha$ fraction: 
$X_{{\rm Ly}\alpha}^{55} < 10${\%} ($0/10$).

\begin{figure*}
   \includegraphics[scale=0.75]{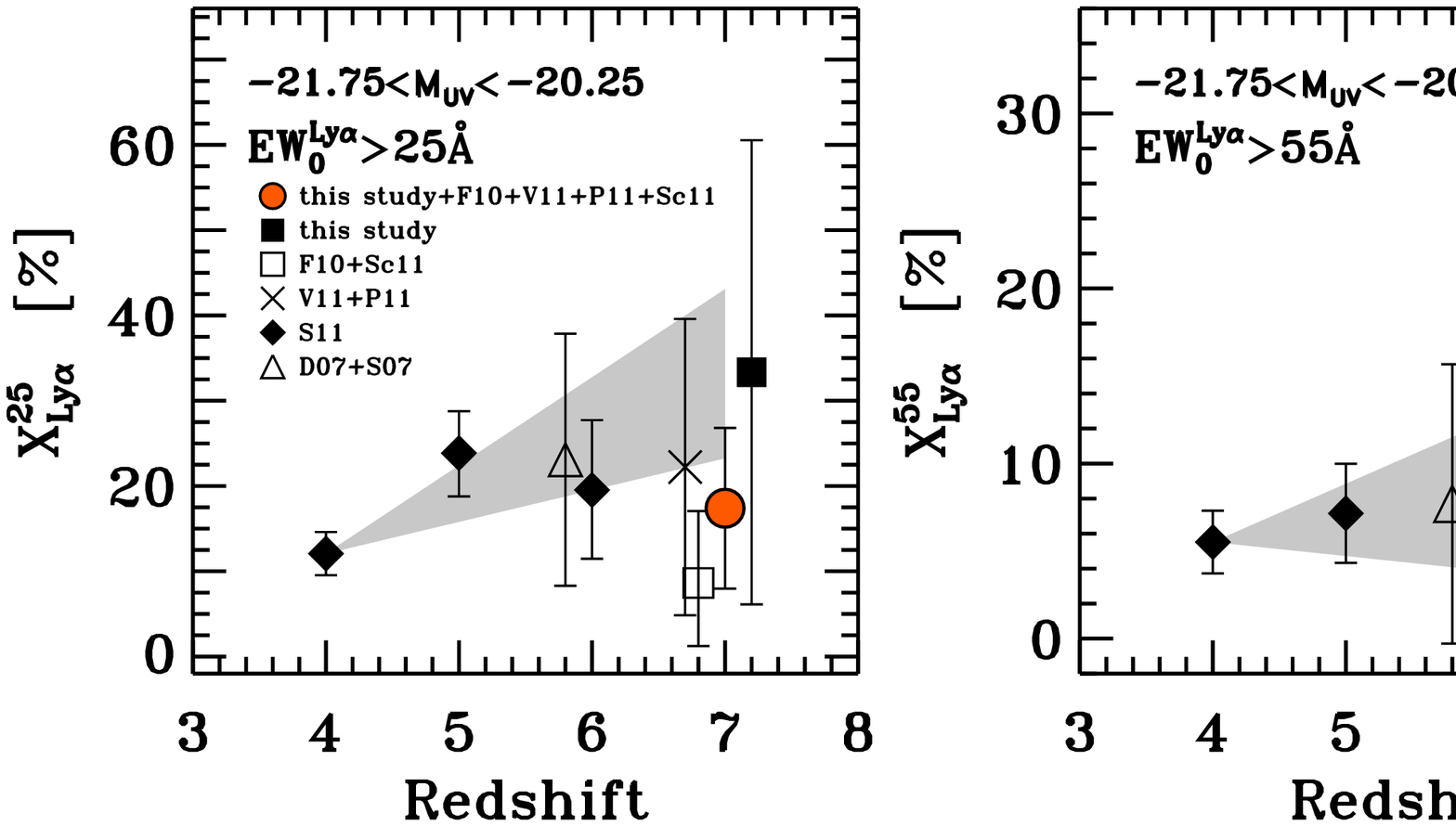}
   \includegraphics[scale=0.75]{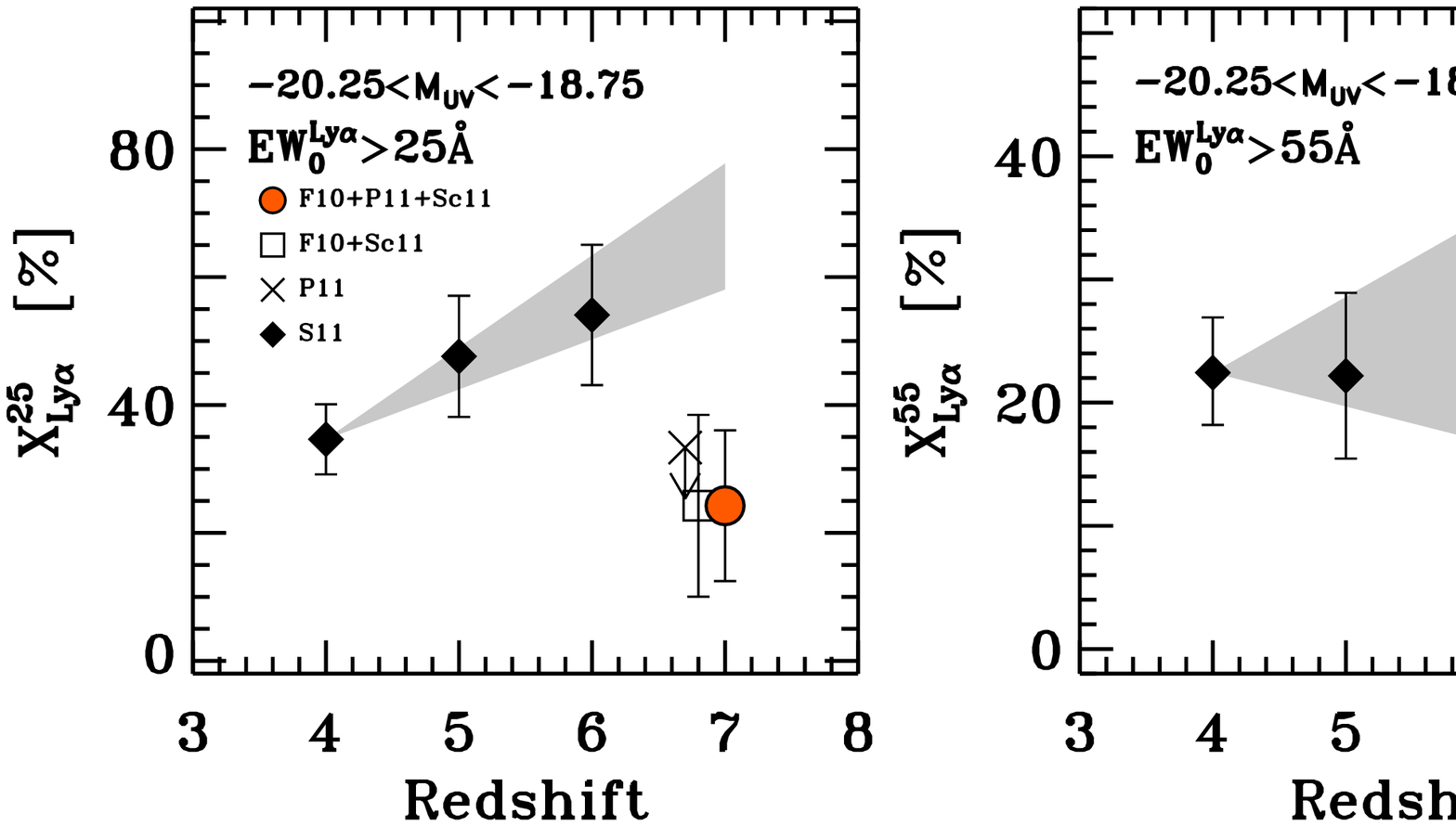}
 \caption[]
{
Evolution in the fraction of strong LAEs in LBGs 
with $-21.75 < M_{\rm UV} <  -20.25$ (top panels) 
and 
$-20.25 < M_{\rm UV} <  -18.75$ (bottom panels) 
over $4 < z < 7$.
The left panels show the fraction of galaxies with EW larger than $25${\AA}, 
while the right panels show the fraction of those with EW larger than $55${\AA}.  
The filled square is our results, 
the open square is the results of \cite{fontana2010} and \cite{schenker2011}, 
the cross is from \cite{vanzella2010d} and \cite{pentericci2011}, 
and the filled circle is the composite results.
The filled diamonds are the results of \cite{stark2010b}, 
and open triangle is the composite result of \cite{dow2007} and \cite{stanway2007}. 
The filled square, open square, cross, and open triangle 
are shifted in redshift for clarity. 
The shaded area is derived by extrapolating the trend seen in lower redshifts to $z \sim 7$ \citep{stark2010b}.  
}
\label{fig:Lyafrac}
\end{figure*}

\section{DISCUSSION: Implications for Reionization} \label{sec:discussion}

Since neutral hydrogen in the IGM resonantly scatters Ly$\alpha$ photons, 
the transmission of Ly$\alpha$ photons is sensitive to the ionization state of the IGM. 
Thus, the fraction of Ly$\alpha$-emitting LBGs can be used 
as a diagnostic for the ionization state of the IGM. 
In Section \ref{subsec:X_Lya}, 
we estimated the Ly$\alpha$ fraction of bright $z$-dropout galaxies 
to be 
$X_{{\rm Ly}\alpha}^{25} = 33 \pm 27${\%} ($2/6$), and 
$X_{{\rm Ly}\alpha}^{55} < 10${\%} ($0/10$).
In order to study evolution of the Ly$\alpha$ fraction to $z \sim 7$, 
we compare our estimates for the bright 
($-21.75 < M_{\rm UV} < -20.25$, or $M_{\rm UV} \simeq -21 $)
$z$-dropout galaxies 
with those at $z \sim (4, \, 5, \, 6)$, 
$X^{25}_{{\rm Ly}\alpha} = (12 \pm 3${\%}, \, $24 \pm 5${\%}, \, $20 \pm 8${\%})    
and 
$X^{55}_{{\rm Ly}\alpha} = (5.5 \pm 1.8${\%}, \, $7.2 \pm 2.8${\%}, \, $7.5 \pm 5.0${\%}) 
and their extrapolations to $z \sim 7$, 
$X^{25}_{{\rm Ly}\alpha} = 33 \pm 10${\%}
and 
$X^{55}_{{\rm Ly}\alpha} = 9 \pm 6${\%}, 
which are derived on the assumption of a linear relationship 
between Ly$\alpha$ fraction and redshift \citep{stark2010b}. 
For faint 
($-20.25 < M_{\rm UV} < -18.75$, or $M_{\rm UV} \simeq -19.5$)
galaxies at $z \sim (4, \, 5, \, 6)$, 
\cite{stark2010b} obtained 
their Ly$\alpha$ fractions of 
$X^{25}_{{\rm Ly}\alpha} = (35 \pm 5${\%}, \, $48 \pm 9${\%}, \, $54 \pm 11${\%})    
and 
$X^{55}_{{\rm Ly}\alpha} = (22 \pm 4${\%}, \, $22 \pm 7${\%}, \, $27 \pm 8${\%}) 
and their extrapolations to $z \sim 7$, 
$X^{25}_{{\rm Ly}\alpha} = 67 \pm 10${\%}
and 
$X^{55}_{{\rm Ly}\alpha} = 27 \pm 13${\%}.

\begin{figure}
   \includegraphics[scale=0.65]{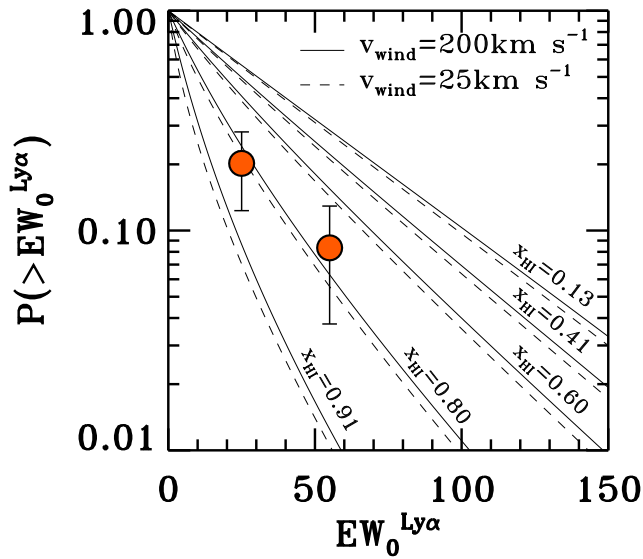}
 \caption[]
{
Cumulative distribution function (CDF) of rest-frame Ly$\alpha$ EW for $z \sim 7$ galaxies.
The filled circles are the composite results.
The solid lines show the $z=7$ CDF 
for the wind model of \cite{dijkstra2011} with 
column density $N_{\rm HI} = 10^{20}$ cm$^{-2}$, 
wind velocity $v_{\rm wind} = 200$ km s$^{-1}$, 
and neutral hydrogen fraction $x_{\rm HI}$ of 
$0.13$, $0.41$, $0.60$, $0.80$, $0.91$ from top to bottom.
The dashed lines are the same as the solid lines, 
except that $v_{\rm wind} = 25$ km s$^{-1}$. 
}
\label{fig:EW_Lyafrac}
\end{figure}

For an independent check 
on the values from \cite{stark2010b}, 
we refer to \cite{dow2007} and \cite{stanway2007}, 
who presented the results of spectroscopy for 
$i$-dropout galaxies at $z \sim 6$. 
\cite{dow2007} reported a total of $12$ LBGs 
with bright UV continuum ($-21.75 < M_{\rm UV} < -20.25$)\footnote{
BD 00, 
BD 03, 
BD 27, 
BD 44, 
BD 46, 
BD 58, 
BD 66, 
GOODS i6 0, 
UDF PFs i0, 
UDF PFs i4, 
UDF PFs IDROP1, 
UDF PFs1 i9  
\citep[Table 4 of][]{dow2007}. }. 
Three (one) of these galaxies are found to have 
Ly$\alpha$ EWs larger than $25${\AA} ($55${\AA}). 
In an independent study, 
\cite{stanway2007} found one such bright galaxy (ID 1042) 
with Ly$\alpha$ EW of $23${\AA}. 
Combining these results, 
we estimate a Ly$\alpha$ fraction of 
$X^{25}_{{\rm Ly}\alpha} = 23 \pm 15${\%} ($3/13$) 
and 
$X^{55}_{{\rm Ly}\alpha} = 8 \pm 8${\%} ($1/13$) 
at $z \sim 6$, 
consistent with \cite{stark2010b}. 
Since they would not change so much the extrapolations 
derived by \cite{stark2010b} due to 
the large statistical uncertainties of the combined results,  
we do not calculate extrapolations using the combined results 
as well as the results of \cite{stark2010b}.

\cite{fontana2010} reported 
spectroscopic observation of seven $z$-dropout galaxies, 
with six
having bright UV continua ($-21.75 < M_{\rm UV} < -20.25$), 
and the other one having 
faint UV continuum ($-20.25 < M_{\rm UV} < -18.75$). 
One of their galaxies shows Ly$\alpha$ emission with EW$_0 = 13${\AA}. 
Furthermore, 
\cite{vanzella2010d} performed spectroscopic observations 
of two $z$-dropout candidates. 
Both of them have $M_{\rm UV} < -20.25$ 
and the two galaxies 
show Ly$\alpha$ emission with EWs of $64${\AA} and $50${\AA}. 
\cite{schenker2011} showed the results of spectroscopic observations 
of $19$ galaxy candidates at $6.3< z < 8.8$ selected by their 
photometric redshift technique \citep{mclure2011}. 
They reported three Ly$\alpha$ detections with EWs of 
$65${\AA}, $65${\AA}, and $30${\AA}. 
\cite{pentericci2011} presented the results of their spectroscopy 
for $11$ $z$-dropout candidates, among which 
one galaxy is considered to be an interloper 
because of its multiple line detection. 
Seven galaxies have bright UV continua, 
while the remaining three galaxies are faint. 
Although two of the $10$ candidates show Ly$\alpha$ detection, 
their Ly$\alpha$ EWs are lower than $25${\AA}.
Note that Ly$\alpha$ EW detection limits for the objects of \cite{pentericci2011} 
is well below EW$_0^{{\rm Ly}\alpha} = 25${\AA}. 
Table \ref{tab:previous} summarizes the spectroscopic results 
of $z$-dropout galaxies 
by \cite{fontana2010}, 
\cite{vanzella2010d}, 
\cite{schenker2011}, 
and \cite{pentericci2011}. 
\cite{schenker2011} estimated 
the Ly$\alpha$ fraction 
based on their results and the results of \cite{fontana2010}, 
correcting the effects of 
instrumentally-limited Ly$\alpha$ visibilities in wavelength 
and strong OH airglow lines. 
They obtained 
$X^{25}_{{\rm Ly}\alpha} = 8.6^{+8.4}_{-7.4}${\%} for UV-bright galaxies and
$X^{25}_{{\rm Ly}\alpha} = 24 \pm 14${\%} for UV-faint galaxies.
We estimate the Ly$\alpha$ fraction of 
$z$-dropout galaxies 
from \cite{vanzella2010d} and \cite{pentericci2011}
to be 
$X^{25}_{{\rm Ly}\alpha} = 22^{+18}_{-17}${\%} ($2/9$) and  
$X^{55}_{{\rm Ly}\alpha} = 11^{+12}_{-11}${\%} ($1/9$) 
for UV-bright galaxies at $z \sim 7$.  
For UV-faint galaxies,
the upper limit of the Ly$\alpha$ fraction is 
$X^{25}_{{\rm Ly}\alpha} < 33${\%} ($0/3$) and  
$X^{55}_{{\rm Ly}\alpha} < 33${\%} ($0/3$). 
Comparing estimates of the Ly$\alpha$ fraction 
for $z \sim 7$ galaxies here, 
we find a spread due to small number statistics 
and possible field-to-field variation. 
Therefore, 
to minimize the Poisson noise and cosmic variance, 
we combine estimates of Ly$\alpha$ fraction 
from these independent studies. 
Combining these estimates assumes that 
the field-to-field variation is due to cosmic variance, 
and is not indicative of patchy reionization.

We now combine our results for $z \sim 7$ galaxies here 
with those from 
\cite{fontana2010}, 
\cite{vanzella2010d}, 
\cite{pentericci2011}, 
and \cite{schenker2011},  
to measure the Ly$\alpha$ fraction 
using the spectra of 
$z$-dropout galaxies 
obtained by three different groups.
For the objects of \cite{schenker2011}, 
we use $14$ objects whose redshifts are 
in the range of $6.3 < z < 7.3$,  
and take into account 
their correction factors for Ly$\alpha$ visibilities. 
Using the combined data, we 
calculate number-weighted mean Ly$\alpha$ fractions:  
$X^{25}_{{\rm Ly}\alpha} = 17^{+10}_{-9}${\%} 
and 
$X^{55}_{{\rm Ly}\alpha} = 4 \pm 4${\%} 
for UV-bright galaxies, 
and
$X^{25}_{{\rm Ly}\alpha} = 24 \pm 12${\%} 
and 
$X^{55}_{{\rm Ly}\alpha} = 16 \pm 9${\%} 
for UV-faint galaxies. 
In Figure \ref{fig:Lyafrac}, 
we present the evolution of Ly$\alpha$ fraction 
with redshift for LBGs over the range $4 < z < 7$. 
We also show, as shaded area, 
an extrapolation of lower-$z$ Ly$\alpha$ fraction 
to $z \sim 7$, 
which is derived by \cite{stark2010b}. 
We find that 
the Ly$\alpha$ fractions $X_{\rm Lya}^{25}$ 
drop from $z \sim 6$ to $7$ 
based on the combined results. 
These findings are similar to 
those reported by \cite{schenker2011} and \cite{pentericci2011}.

\cite{stark2010} reported that 
Ly$\alpha$ fraction increases with redshift over $3 < z < 6$
at fixed luminosity, 
and the trend is likely governed by redshift-dependent variations 
in dust obscuration, 
with additional contributions from 
kinematics and covering fraction of neutral hydrogen.  
In contrast to this increasing tendency
of Ly$\alpha$ fraction from $z \sim 3$ to $6$, 
we have found a significant drop of Ly$\alpha$ fraction 
from $z \sim 6$ to $7$. 
This significant drop would suggest that 
Ly$\alpha$ emission lines from $z \sim 7$ galaxies
are scattered by the IGM 
whose neutral hydrogen fraction is higher than that at $z \sim 6$.
We would witness the increase of neutral hydrogen fraction
toward $z \sim 7$ in the cosmic reionization history.

In Figure \ref{fig:Lyafrac}, 
we find that $X_{\rm Lya}^{25}$ drops more strongly 
in the faint ($M_{\rm UV} \simeq -19.5$) galaxies 
than in the bright ($M_{\rm UV} \simeq -21$) galaxies; 
$X_{\rm Lya}^{25} (z=7; \, {\rm obs}) \, / \, X_{\rm Lya}^{25} (z=7; \, {\rm exp}) = 0.36 \pm 0.18$, and $0.53 \pm 0.33$,
respectively, 
where $X_{\rm Lya}^{25} (z=7; {\rm obs})$ is 
the observed Ly$\alpha$ fraction at $z=7$, 
and $X_{\rm Lya}^{25} (z=7; {\rm exp})$ is 
the expected Ly$\alpha$ fraction at $z=7$ derived by extrapolating the trend seen in lower redshifts to $z \sim 7$.  
This magnitude dependence of $X_{\rm Lya}^{25}$ evolution 
could be explained by 
different halo masses of galaxies and the surrounding IGM.
Given that 
the clustering strength of dropout galaxies increases 
with their UV luminosity \citep[e.g.,][]{giavalisco2001,ouchi2004b,adelberger2005,leek2006}, 
our results imply that the ionizing state of the IGM 
around galaxies hosted by less-massive dark matter halos 
changes later than 
that around galaxies hosted by massive dark matter halos.
This would suggest that 
reionization proceeds from high- to low-density environments 
\citep[inside-out; e.g.,][c.f., \citealt{finlator2009b}]{ciardi2003,sokasian2002,iliev2006}
rather than from low- to high-density regions 
\citep[outside-in; e.g.,][]{gnedin2000,miralda2000}.

We compare our composite results with those of model predictions 
derived by \cite{dijkstra2011}, 
which quantify the probability distribution function (PDF) of 
the fraction of Ly$\alpha$ photons transmitted through the IGM, 
by combining galactic outflow models 
with large-scale seminumeric simulations of reionization. 
They assume that the IGM at $z=6$ was fully transparent to Ly$\alpha$ photons, 
and that the observed PDF for EW$_0^{{\rm Ly}\alpha}$ at $z=7$ is different 
only because of evolution of the ionization state of the IGM.
Figure \ref{fig:EW_Lyafrac} compares their models 
with our composite results.  
Our results can be explained by an evolution of the neutral hydrogen fraction $x_{\rm HI}$
between $z=6$ and $7$; 
$x_{\rm HI}$ is roughly $0.6-0.9$ at $z \sim 7$, 
which is similar to those reported by \cite{schenker2011} and \cite{pentericci2011}.

The above discussion assumes that 
the dropout samples at different redshifts have similarly 
low contamination fractions from interlopers. 
We should, however, keep in mind that this assumption has not 
yet been justified well, although 
for the $z = 4-6$ UV-faint samples, \cite{stark2010} estimated 
the contamination rate to be only $2-5$ percent \citep[c.f.,][]{stanway2008,douglas2009}.
Another possible source of systematic errors 
in our Ly$\alpha$ fraction analysis is 
inhomogeneities among the dropout samples 
in the spectroscopic detection limit and in the quality of 
the photometry used to estimate UV continua, 
both of which are difficult to fully take into account 
in our analysis.
To reduce such inhomogeneities, desirable is 
a systematic spectroscopic survey over an entire redshift range 
combined with deep photometry redward of Ly$\alpha$ wavelength.

\section{CONCLUSIONS} \label{sec:conclusion}

In this paper, 
we have presented Keck/DEIMOS spectroscopic observations
 of $11$ $z$-dropout galaxies 
found in the SDF and GOODS-N fields. 
An emission line
has been detected 
at $9500 - 10000${\AA} 
in the spectra of three objects,  
one of which is the previously reported 
Ly$\alpha$ emitter IOK-1 at $z = 6.96$. 
Since all the detected lines are singlet 
with a large positive weighted skewness, 
we have concluded that 
the three objects are Ly$\alpha$-emitting $z$-dropout galaxies 
at $z_{\rm spec} = 7.213$, $6.965$, and $6.844$.  
Their Ly$\alpha$ fluxes and rest-frame Ly$\alpha$ EWs are 
$2.5 \times 10^{-17}$ erg s$^{-1}$ cm$^{-2}$ and $33${\AA}, 
$2.8 \times 10^{-17}$ erg s$^{-1}$ cm$^{-2}$ and $43${\AA}, 
$2.7 \times 10^{-17}$ erg s$^{-1}$ cm$^{-2}$ and $43${\AA}, respectively. 
It should be noted that 
the $z=7.213$ galaxy 
was confirmed by observations in two independent DEIMOS runs in $2010$ and $2011$ 
with three different spectroscopic configurations. 
This galaxy is detected 
in HST/WFC3 F140W imaging  
as well as in Spitzer/IRAC $3.6$ and 4.5 $\mu$m imaging. 
Stellar population modeling indicates that 
this galaxy has a very young age, 
a stellar mass $\lesssim 10^9 M_\odot$, 
and a high star formation rate of $30 -100 M_\odot$ yr$^{-1}$, 
with strong nebular emission 
contributing to the fluxes in the IRAC bands.

We then have measured the fraction of 
Ly$\alpha$-emitting galaxies at $z \sim 7$. 
To reduce statistical uncertainties 
and possible effects of field-to-field variance, 
we have combined our results 
with previous $z$-dropout spectroscopic studies \citep{fontana2010,vanzella2010d} 
including very recent ones \citep{schenker2011,pentericci2011}.  
We have obtained the $z \sim 7$ Ly$\alpha$ fraction of 
$X^{25}_{{\rm Ly}\alpha} = 17 \pm 10${\%}, and  
$X^{55}_{{\rm Ly}\alpha} = 4 \pm 4${\%} for UV-bright galaxies 
($-21.75 < M_{\rm UV} < -20.25$, or $M_{\rm UV} \simeq -21$), 
and 
$X^{25}_{{\rm Ly}\alpha} = 24 \pm 12${\%}, and  
$X^{55}_{{\rm Ly}\alpha} = 16 \pm 9${\%} for UV-faint galaxies 
($-20.25 < M_{\rm UV} < -18.75$, or $M_{\rm UV} \simeq -19.5$). 
These low values indicate that 
the fraction of Ly$\alpha$-emitting galaxies 
drops from $z \sim 6$ to $7$, 
in contrast to the reported increasing trend 
from $z \sim 4$ to $6$. 
We have also found that 
$X^{25}_{{\rm Ly}\alpha}$ drops more strongly 
in UV-faint galaxies than in UV-bright galaxies. 
These findings would suggest that 
the neutral fraction of the IGM 
significantly increases from $z \sim 6$ to $7$, 
and that 
the increase is stronger around galaxies 
with fainter UV luminosities, 
which is consistent with 
inside-out reionization models 
where reionization proceeds 
from high- to low-density environments.

\section*{Acknowledgements}

We thank the anonymous referee for valuable comments and suggestions 
which improved the manuscript.
The authors wish to recognize and acknowledge
the very significant cultural role and reverence
that the summit of Mauna Kea has always had
within the indigenous Hawaiian community.
We are most fortunate to have the opportunity
to conduct observations from this mountain.
We would like to thank Michael Cooper and Yousuke Utsumi 
for giving us helpful advices to reduce DEIMOS spectra. 
We are also thankful to Richard Ellis 
for giving us helpful comments on an early draft of this paper, 
and to Mark Dijkstra 
for providing us with the machine-readable table of their simulation results.
Y.O. acknowledges support 
from the Japan Society for the Promotion of Science (JSPS) 
through the JSPS Research Fellowship for Young Scientists.
The work of DS was carried
out at Jet Propulsion Laboratory, California Institute of Technology,
under a contract with NASA.
H.S. would like to acknowledge the support of
the National Science Foundation during the earlier phases of
his spectroscopic program at the Keck Observatory.

Facilities: Subaru (Suprime-Cam), Keck (DEIMOS).


\bibliographystyle{apj}
\bibliography{apj-jour,../../../Papers/papers_gal/ms}

\begin{thebibliography}{111}
\expandafter\ifx\csname natexlab\endcsname\relax\def\natexlab#1{#1}\fi

\bibitem[{{Adelberger} {et~al.}(2005){Adelberger}, {Steidel}, {Pettini},
  {Shapley}, {Reddy}, \& {Erb}}]{adelberger2005}
{Adelberger}, K.~L., {Steidel}, C.~C., {Pettini}, M., {Shapley}, A.~E.,
  {Reddy}, N.~A., \& {Erb}, D.~K. 2005, \apj, 619, 697

\bibitem[{{Atek} {et~al.}(2010){Atek}, {Malkan}, {McCarthy}, {Teplitz},
  {Scarlata}, {Siana}, {Henry}, {Colbert}, {Ross}, {Bridge}, {Bunker},
  {Dressler}, {Fosbury}, {Martin}, \& {Shim}}]{atek2010}
{Atek}, H., {et~al.} 2010, \apj, 723, 104

\bibitem[{{Bouwens} {et~al.}(2010{\natexlab{a}}){Bouwens}, {Illingworth},
  {Oesch}, {Stiavelli}, {van Dokkum}, {Trenti}, {Magee}, {Labb{\'e}}, {Franx},
  {Carollo}, \& {Gonzalez}}]{bouwens2010}
{Bouwens}, R.~J., {et~al.} 2010{\natexlab{a}}, \apjl, 709, L133

\bibitem[{{Bouwens} {et~al.}(2010{\natexlab{b}}){Bouwens}, {Illingworth},
  {Oesch}, {Trenti}, {Stiavelli}, {Carollo}, {Franx}, {van Dokkum},
  {Labb{\'e}}, \& {Magee}}]{bouwens2010b}
---. 2010{\natexlab{b}}, \apjl, 708, L69

\bibitem[{{Bouwens} {et~al.}(2011{\natexlab{a}}){Bouwens}, {Illingworth},
  {Labbe}, {Oesch}, {Trenti}, {Carollo}, {van Dokkum}, {Franx}, {Stiavelli},
  {Gonz{\'a}lez}, {Magee}, \& {Bradley}}]{bouwens2011b}
---. 2011{\natexlab{a}}, \nat, 469, 504

\bibitem[{{Bouwens} {et~al.}(2011{\natexlab{b}}){Bouwens}, {Illingworth},
  {Oesch}, {Labb{\'e}}, {Trenti}, {van Dokkum}, {Franx}, {Stiavelli},
  {Carollo}, {Magee}, \& {Gonzalez}}]{bouwens2011}
---. 2011{\natexlab{b}}, \apj, 737, 90

\bibitem[{{Bradley} {et~al.}(2011){Bradley}, {Bouwens}, {Zitrin}, {Smit},
  {Coe}, {Ford}, {Zheng}, {Illingworth}, {Ben{\'{\i}}tez}, \&
  {Broadhurst}}]{bradley2011}
{Bradley}, L.~D., {et~al.} 2011, ArXiv e-prints (arXiv:1104.2035)

\bibitem[{{Bruzual} \& {Charlot}(2003)}]{bc2003}
{Bruzual}, G., \& {Charlot}, S. 2003, \mnras, 344, 1000

\bibitem[{{Bunker} {et~al.}(2010){Bunker}, {Wilkins}, {Ellis}, {Stark},
  {Lorenzoni}, {Chiu}, {Lacy}, {Jarvis}, \& {Hickey}}]{bunker2010}
{Bunker}, A.~J., {et~al.} 2010, \mnras, 409, 855

\bibitem[{{Cai} {et~al.}(2011){Cai}, {Fan}, {Jiang}, {Bian}, {McGreer},
  {Dav{\'e}}, {Egami}, {Zabludoff}, {Yang}, \& {Oh}}]{cai2011}
{Cai}, Z., {et~al.} 2011, \apjl, 736, L28+

\bibitem[{{Calzetti} {et~al.}(2000){Calzetti}, {Armus}, {Bohlin}, {Kinney},
  {Koornneef}, \& {Storchi-Bergmann}}]{calzetti2000}
{Calzetti}, D., {Armus}, L., {Bohlin}, R.~C., {Kinney}, A.~L., {Koornneef}, J.,
  \& {Storchi-Bergmann}, T. 2000, \apj, 533, 682

\bibitem[{{Castellano} {et~al.}(2010{\natexlab{a}}){Castellano}, {Fontana},
  {Boutsia}, {Grazian}, {Pentericci}, {Bouwens}, {Dickinson}, {Giavalisco},
  {Santini}, {Cristiani}, {Fiore}, {Gallozzi}, {Giallongo}, {Maiolino},
  {Mannucci}, {Menci}, {Moorwood}, {Nonino}, {Paris}, {Renzini}, {Rosati},
  {Salimbeni}, {Testa}, \& {Vanzella}}]{castellano2010}
{Castellano}, M., {et~al.} 2010{\natexlab{a}}, \aap, 511, A20+

\bibitem[{{Castellano} {et~al.}(2010{\natexlab{b}}){Castellano}, {Fontana},
  {Paris}, {Grazian}, {Pentericci}, {Boutsia}, {Santini}, {Testa}, {Dickinson},
  {Giavalisco}, {Bouwens}, {Cuby}, {Mannucci}, {Cl{\'e}ment}, {Cristiani},
  {Fiore}, {Gallozzi}, {Giallongo}, {Maiolino}, {Menci}, {Moorwood}, {Nonino},
  {Renzini}, {Rosati}, {Salimbeni}, \& {Vanzella}}]{castellano2010b}
---. 2010{\natexlab{b}}, \aap, 524, A28+

\bibitem[{{Ciardi} \& {Madau}(2003)}]{ciardi2003}
{Ciardi}, B., \& {Madau}, P. 2003, \apj, 596, 1

\bibitem[{{Davis} {et~al.}(2003){Davis}, {Faber}, {Newman}, {Phillips},
  {Ellis}, {Steidel}, {Conselice}, {Coil}, {Finkbeiner}, {Koo}, {Guhathakurta},
  {Weiner}, {Schiavon}, {Willmer}, {Kaiser}, {Luppino}, {Wirth}, {Connolly},
  {Eisenhardt}, {Cooper}, \& {Gerke}}]{davis2003}
{Davis}, M., {et~al.} 2003, in Society of Photo-Optical Instrumentation
  Engineers (SPIE) Conference Series, Vol. 4834, Society of Photo-Optical
  Instrumentation Engineers (SPIE) Conference Series, ed. {P.~Guhathakurta},
  161--172

\bibitem[{{Dayal} {et~al.}(2008){Dayal}, {Ferrara}, \& {Gallerani}}]{dayal2008}
{Dayal}, P., {Ferrara}, A., \& {Gallerani}, S. 2008, \mnras, 389, 1683

\bibitem[{{Dayal} {et~al.}(2009){Dayal}, {Ferrara}, {Saro}, {Salvaterra},
  {Borgani}, \& {Tornatore}}]{dayal2009}
{Dayal}, P., {Ferrara}, A., {Saro}, A., {Salvaterra}, R., {Borgani}, S., \&
  {Tornatore}, L. 2009, \mnras, 400, 2000

\bibitem[{{Dayal} {et~al.}(2011){Dayal}, {Maselli}, \& {Ferrara}}]{dayal2011}
{Dayal}, P., {Maselli}, A., \& {Ferrara}, A. 2011, \mnras, 410, 830

\bibitem[{{Dijkstra} {et~al.}(2007){Dijkstra}, {Lidz}, \&
  {Wyithe}}]{dijkstra2007}
{Dijkstra}, M., {Lidz}, A., \& {Wyithe}, J.~S.~B. 2007, \mnras, 377, 1175

\bibitem[{{Dijkstra} {et~al.}(2011){Dijkstra}, {Mesinger}, \&
  {Wyithe}}]{dijkstra2011}
{Dijkstra}, M., {Mesinger}, A., \& {Wyithe}, J.~S.~B. 2011, \mnras, 414, 2139

\bibitem[{{Douglas} {et~al.}(2010){Douglas}, {Bremer}, {Lehnert}, {Stanway}, \&
  {Milvang-Jensen}}]{douglas2010}
{Douglas}, L.~S., {Bremer}, M.~N., {Lehnert}, M.~D., {Stanway}, E.~R., \&
  {Milvang-Jensen}, B. 2010, \mnras, 409, 1155

\bibitem[{{Douglas} {et~al.}(2009){Douglas}, {Bremer}, {Stanway}, {Lehnert}, \&
  {Clowe}}]{douglas2009}
{Douglas}, L.~S., {Bremer}, M.~N., {Stanway}, E.~R., {Lehnert}, M.~D., \&
  {Clowe}, D. 2009, \mnras, 400, 561

\bibitem[{{Dow-Hygelund} {et~al.}(2007){Dow-Hygelund}, {Holden}, {Bouwens},
  {Illingworth}, {van der Wel}, {Franx}, {van Dokkum}, {Ford}, {Rosati},
  {Magee}, \& {Zirm}}]{dow2007}
{Dow-Hygelund}, C.~C., {et~al.} 2007, \apj, 660, 47

\bibitem[{{Dunkley} {et~al.}(2009){Dunkley}, {Komatsu}, {Nolta}, {Spergel},
  {Larson}, {Hinshaw}, {Page}, {Bennett}, {Gold}, {Jarosik}, {Weiland},
  {Halpern}, {Hill}, {Kogut}, {Limon}, {Meyer}, {Tucker}, {Wollack}, \&
  {Wright}}]{dunkley2009}
{Dunkley}, J., {et~al.} 2009, \apjs, 180, 306

\bibitem[{{Dunlop} {et~al.}(2011){Dunlop}, {McLure}, {Robertson}, {Ellis},
  {Stark}, {Cirasuolo}, \& {de Ravel}}]{dunlop2011}
{Dunlop}, J.~S., {McLure}, R.~J., {Robertson}, B.~E., {Ellis}, R.~S., {Stark},
  D.~P., {Cirasuolo}, M., \& {de Ravel}, L. 2011, ArXiv e-prints
  (arXiv:1102.5005)

\bibitem[{{Erb} {et~al.}(2006){Erb}, {Steidel}, {Shapley}, {Pettini}, {Reddy},
  \& {Adelberger}}]{erb2006c}
{Erb}, D.~K., {Steidel}, C.~C., {Shapley}, A.~E., {Pettini}, M., {Reddy},
  N.~A., \& {Adelberger}, K.~L. 2006, \apj, 647, 128

\bibitem[{{Faber} {et~al.}(2003){Faber}, {Phillips}, {Kibrick}, {Alcott},
  {Allen}, {Burrous}, {Cantrall}, {Clarke}, {Coil}, {Cowley}, {Davis}, {Deich},
  {Dietsch}, {Gilmore}, {Harper}, {Hilyard}, {Lewis}, {McVeigh}, {Newman},
  {Osborne}, {Schiavon}, {Stover}, {Tucker}, {Wallace}, {Wei}, {Wirth}, \&
  {Wright}}]{faber2003}
{Faber}, S.~M., {et~al.} 2003, in Presented at the Society of Photo-Optical
  Instrumentation Engineers (SPIE) Conference, Vol. 4841, Society of
  Photo-Optical Instrumentation Engineers (SPIE) Conference Series, ed. {M.~Iye
  \& A.~F.~M.~Moorwood}, 1657--1669

\bibitem[{{Fan} {et~al.}(2006){Fan}, {Strauss}, {Becker}, {White}, {Gunn},
  {Knapp}, {Richards}, {Schneider}, {Brinkmann}, \& {Fukugita}}]{fan2006}
{Fan}, X., {et~al.} 2006, \aj, 132, 117

\bibitem[{{Finkelstein} {et~al.}(2010){Finkelstein}, {Papovich}, {Giavalisco},
  {Reddy}, {Ferguson}, {Koekemoer}, \& {Dickinson}}]{finkelstein2009f}
{Finkelstein}, S.~L., {Papovich}, C., {Giavalisco}, M., {Reddy}, N.~A.,
  {Ferguson}, H.~C., {Koekemoer}, A.~M., \& {Dickinson}, M. 2010, \apj, 719,
  1250

\bibitem[{{Finlator} {et~al.}(2009){Finlator}, {{\"O}zel}, {Dav{\'e}}, \&
  {Oppenheimer}}]{finlator2009b}
{Finlator}, K., {{\"O}zel}, F., {Dav{\'e}}, R., \& {Oppenheimer}, B.~D. 2009,
  \mnras, 400, 1049

\bibitem[{{Fontana} {et~al.}(2010){Fontana}, {Vanzella}, {Pentericci},
  {Castellano}, {Giavalisco}, {Grazian}, {Boutsia}, {Cristiani}, {Dickinson},
  {Giallongo}, {Maiolino}, {Moorwood}, \& {Santini}}]{fontana2010}
{Fontana}, A., {et~al.} 2010, \apjl, 725, L205

\bibitem[{{Furlanetto} {et~al.}(2006){Furlanetto}, {Zaldarriaga}, \&
  {Hernquist}}]{furlanetto2006}
{Furlanetto}, S.~R., {Zaldarriaga}, M., \& {Hernquist}, L. 2006, \mnras, 365,
  1012

\bibitem[{{Giavalisco}(2002)}]{giavalisco2002}
{Giavalisco}, M. 2002, \araa, 40, 579

\bibitem[{{Giavalisco} \& {Dickinson}(2001)}]{giavalisco2001}
{Giavalisco}, M., \& {Dickinson}, M. 2001, \apj, 550, 177

\bibitem[{{Giavalisco} {et~al.}(2004){Giavalisco}, {Ferguson}, {Koekemoer},
  {Dickinson}, {Alexander}, {Bauer}, {Bergeron}, {Biagetti}, {Brandt},
  {Casertano}, {Cesarsky}, {Chatzichristou}, {Conselice}, {Cristiani}, {Da
  Costa}, {Dahlen}, {de Mello}, {Eisenhardt}, {Erben}, {Fall}, {Fassnacht},
  {Fosbury}, {Fruchter}, {Gardner}, {Grogin}, {Hook}, {Hornschemeier}, {Idzi},
  {Jogee}, {Kretchmer}, {Laidler}, {Lee}, {Livio}, {Lucas}, {Madau},
  {Mobasher}, {Moustakas}, {Nonino}, {Padovani}, {Papovich}, {Park},
  {Ravindranath}, {Renzini}, {Richardson}, {Riess}, {Rosati}, {Schirmer},
  {Schreier}, {Somerville}, {Spinrad}, {Stern}, {Stiavelli}, {Strolger},
  {Urry}, {Vandame}, {Williams}, \& {Wolf}}]{giavalisco2004}
{Giavalisco}, M., {et~al.} 2004, \apjl, 600, L93

\bibitem[{{Gnedin}(2000)}]{gnedin2000}
{Gnedin}, N.~Y. 2000, \apj, 535, 530

\bibitem[{{Gonz{\'a}lez} {et~al.}(2011){Gonz{\'a}lez}, {Labb{\'e}}, {Bouwens},
  {Illingworth}, {Franx}, \& {Kriek}}]{gonzalez2011}
{Gonz{\'a}lez}, V., {Labb{\'e}}, I., {Bouwens}, R.~J., {Illingworth}, G.,
  {Franx}, M., \& {Kriek}, M. 2011, \apjl, 735, L34+

\bibitem[{{Gonz{\'a}lez} {et~al.}(2010){Gonz{\'a}lez}, {Labb{\'e}}, {Bouwens},
  {Illingworth}, {Franx}, {Kriek}, \& {Brammer}}]{gonzalez2009}
{Gonz{\'a}lez}, V., {Labb{\'e}}, I., {Bouwens}, R.~J., {Illingworth}, G.,
  {Franx}, M., {Kriek}, M., \& {Brammer}, G.~B. 2010, \apj, 713, 115

\bibitem[{{Goto} {et~al.}(2011){Goto}, {Utsumi}, {Hattori}, {Miyazaki}, \&
  {Yamauchi}}]{goto2011c}
{Goto}, T., {Utsumi}, Y., {Hattori}, T., {Miyazaki}, S., \& {Yamauchi}, C.
  2011, \mnras, 415, L1

\bibitem[{{Grazian} {et~al.}(2011){Grazian}, {Castellano}, {Koekemoer},
  {Fontana}, {Pentericci}, {Testa}, {Boutsia}, {Giallongo}, {Giavalisco}, \&
  {Santini}}]{grazian2011}
{Grazian}, A., {et~al.} 2011, \aap, 532, A33+

\bibitem[{{Haiman} \& {Cen}(2005)}]{haiman2005}
{Haiman}, Z., \& {Cen}, R. 2005, \apj, 623, 627

\bibitem[{{Haiman} \& {Spaans}(1999)}]{haiman1999}
{Haiman}, Z., \& {Spaans}, M. 1999, \apj, 518, 138

\bibitem[{{Hickey} {et~al.}(2010){Hickey}, {Bunker}, {Jarvis}, {Chiu}, \&
  {Bonfield}}]{hickey2010}
{Hickey}, S., {Bunker}, A., {Jarvis}, M.~J., {Chiu}, K., \& {Bonfield}, D.
  2010, \mnras, 404, 212

\bibitem[{{Hu} {et~al.}(2004){Hu}, {Cowie}, {Capak}, {McMahon}, {Hayashino}, \&
  {Komiyama}}]{hu2004}
{Hu}, E.~M., {Cowie}, L.~L., {Capak}, P., {McMahon}, R.~G., {Hayashino}, T., \&
  {Komiyama}, Y. 2004, \aj, 127, 563

\bibitem[{{Hu} {et~al.}(1998){Hu}, {Cowie}, \& {McMahon}}]{hu1998}
{Hu}, E.~M., {Cowie}, L.~L., \& {McMahon}, R.~G. 1998, \apjl, 502, L99+

\bibitem[{{Iliev} {et~al.}(2006){Iliev}, {Mellema}, {Pen}, {Merz}, {Shapiro},
  \& {Alvarez}}]{iliev2006}
{Iliev}, I.~T., {Mellema}, G., {Pen}, U.-L., {Merz}, H., {Shapiro}, P.~R., \&
  {Alvarez}, M.~A. 2006, \mnras, 369, 1625

\bibitem[{{Iliev} {et~al.}(2008){Iliev}, {Shapiro}, {McDonald}, {Mellema}, \&
  {Pen}}]{iliev2008}
{Iliev}, I.~T., {Shapiro}, P.~R., {McDonald}, P., {Mellema}, G., \& {Pen}, U.
  2008, \mnras, 391, 63

\bibitem[{{Iye} {et~al.}(2006){Iye}, {Ota}, {Kashikawa}, {Furusawa},
  {Hashimoto}, {Hattori}, {Matsuda}, {Morokuma}, {Ouchi}, \&
  {Shimasaku}}]{iye2006}
{Iye}, M., {et~al.} 2006, \nat, 443, 186

\bibitem[{{Kajisawa} {et~al.}(2011){Kajisawa}, {Ichikawa}, {Tanaka}, {Yamada},
  {Akiyama}, {Suzuki}, {Tokoku}, {Katsuno Uchimoto}, {Konishi}, {Yoshikawa},
  {Nishimura}, {Omata}, {Ouchi}, {Iwata}, {Hamana}, \&
  {Onodera}}]{kajisawa2011}
{Kajisawa}, M., {et~al.} 2011, \pasj, 63, 379

\bibitem[{{Kakazu} {et~al.}(2007){Kakazu}, {Cowie}, \& {Hu}}]{kakazu2007}
{Kakazu}, Y., {Cowie}, L.~L., \& {Hu}, E.~M. 2007, \apj, 668, 853

\bibitem[{{Kashikawa} {et~al.}(2004){Kashikawa}, {Shimasaku}, {Yasuda},
  {Ajiki}, {Akiyama}, {Ando}, {Aoki}, {Doi}, {Fujita}, {Furusawa}, {Hayashino},
  {Iwamuro}, {Iye}, {Karoji}, {Kobayashi}, {Kodaira}, {Kodama}, {Komiyama},
  {Matsuda}, {Miyazaki}, {Mizumoto}, {Morokuma}, {Motohara}, {Murayama},
  {Nagao}, {Nariai}, {Ohta}, {Okamura}, {Ouchi}, {Sasaki}, {Sato}, {Sekiguchi},
  {Shioya}, {Tamura}, {Taniguchi}, {Umemura}, {Yamada}, \&
  {Yoshida}}]{kashikawa2004}
{Kashikawa}, N., {et~al.} 2004, \pasj, 56, 1011

\bibitem[{{Kashikawa} {et~al.}(2006){Kashikawa}, {Shimasaku}, {Malkan}, {Doi},
  {Matsuda}, {Ouchi}, {Taniguchi}, {Ly}, {Nagao}, {Iye}, {Motohara},
  {Murayama}, {Murozono}, {Nariai}, {Ohta}, {Okamura}, {Sasaki}, {Shioya}, \&
  {Umemura}}]{kashikawa2006}
---. 2006, \apj, 648, 7

\bibitem[{{Kashikawa} {et~al.}(2011){Kashikawa}, {Shimasaku}, {Matsuda},
  {Egami}, {Jiang}, {Nagao}, {Ouchi}, {Malkan}, {Hattori}, {Ota}, {Taniguchi},
  {Okamura}, {Ly}, {Iye}, {Furusawa}, {Shioya}, {Shibuya}, {Ishizaki}, \&
  {Toshikawa}}]{kashikawa2011}
---. 2011, \apj, 734, 119

\bibitem[{{Kennicutt}(1998)}]{kennicutt1998}
{Kennicutt}, Jr., R.~C. 1998, \araa, 36, 189

\bibitem[{{Kobayashi} {et~al.}(2007){Kobayashi}, {Totani}, \&
  {Nagashima}}]{kobayashi2007}
{Kobayashi}, M.~A.~R., {Totani}, T., \& {Nagashima}, M. 2007, \apj, 670, 919

\bibitem[{{Komatsu} {et~al.}(2011){Komatsu}, {Smith}, {Dunkley}, {Bennett},
  {Gold}, {Hinshaw}, {Jarosik}, {Larson}, {Nolta}, {Page}, {Spergel},
  {Halpern}, {Hill}, {Kogut}, {Limon}, {Meyer}, {Odegard}, {Tucker}, {Weiland},
  {Wollack}, \& {Wright}}]{komatsu2010}
{Komatsu}, E., {et~al.} 2011, \apjs, 192, 18

\bibitem[{{Labb{\'e}} {et~al.}(2010){Labb{\'e}}, {Gonz{\'a}lez}, {Bouwens},
  {Illingworth}, {Oesch}, {van Dokkum}, {Carollo}, {Franx}, {Stiavelli},
  {Trenti}, {Magee}, \& {Kriek}}]{labbe2010}
{Labb{\'e}}, I., {et~al.} 2010, \apjl, 708, L26

\bibitem[{{Larson} {et~al.}(2011){Larson}, {Dunkley}, {Hinshaw}, {Komatsu},
  {Nolta}, {Bennett}, {Gold}, {Halpern}, {Hill}, {Jarosik}, {Kogut}, {Limon},
  {Meyer}, {Odegard}, {Page}, {Smith}, {Spergel}, {Tucker}, {Weiland},
  {Wollack}, \& {Wright}}]{larson2011}
{Larson}, D., {et~al.} 2011, \apjs, 192, 16

\bibitem[{{Lee} {et~al.}(2006){Lee}, {Giavalisco}, {Gnedin}, {Somerville},
  {Ferguson}, {Dickinson}, \& {Ouchi}}]{leek2006}
{Lee}, K.-S., {Giavalisco}, M., {Gnedin}, O.~Y., {Somerville}, R.~S.,
  {Ferguson}, H.~C., {Dickinson}, M., \& {Ouchi}, M. 2006, \apj, 642, 63

\bibitem[{{Lehnert} {et~al.}(2010){Lehnert}, {Nesvadba}, {Cuby}, {Swinbank},
  {Morris}, {Cl{\'e}ment}, {Evans}, {Bremer}, \& {Basa}}]{lehnert2010}
{Lehnert}, M.~D., {et~al.} 2010, \nat, 467, 940

\bibitem[{{Lorenzoni} {et~al.}(2011){Lorenzoni}, {Bunker}, {Wilkins},
  {Stanway}, {Jarvis}, \& {Caruana}}]{lorenzoni2010}
{Lorenzoni}, S., {Bunker}, A.~J., {Wilkins}, S.~M., {Stanway}, E.~R., {Jarvis},
  M.~J., \& {Caruana}, J. 2011, \mnras, 414, 1455

\bibitem[{{Madau}(1995)}]{madau1995}
{Madau}, P. 1995, \apj, 441, 18

\bibitem[{{Madau} {et~al.}(1998){Madau}, {Pozzetti}, \&
  {Dickinson}}]{madau1998}
{Madau}, P., {Pozzetti}, L., \& {Dickinson}, M. 1998, \apj, 498, 106

\bibitem[{{Malhotra} \& {Rhoads}(2004)}]{malhotra2004}
{Malhotra}, S., \& {Rhoads}, J.~E. 2004, \apjl, 617, L5

\bibitem[{{Markwardt}(2009)}]{markwardt2009}
{Markwardt}, C.~B. 2009, in Astronomical Society of the Pacific Conference
  Series, Vol. 411, Astronomical Society of the Pacific Conference Series, ed.
  {D.~A.~Bohlender, D.~Durand, \& P.~Dowler}, 251--+

\bibitem[{{McCarthy}(1993)}]{mccarthy1993}
{McCarthy}, P.~J. 1993, \araa, 31, 639

\bibitem[{{McLure} {et~al.}(2010){McLure}, {Dunlop}, {Cirasuolo}, {Koekemoer},
  {Sabbi}, {Stark}, {Targett}, \& {Ellis}}]{mclure2009}
{McLure}, R.~J., {Dunlop}, J.~S., {Cirasuolo}, M., {Koekemoer}, A.~M., {Sabbi},
  E., {Stark}, D.~P., {Targett}, T.~A., \& {Ellis}, R.~S. 2010, \mnras, 403,
  960

\bibitem[{{McLure} {et~al.}(2011){McLure}, {Dunlop}, {de Ravel}, {Cirasuolo},
  {Ellis}, {Schenker}, {Robertson}, {Koekemoer}, {Stark}, \&
  {Bowler}}]{mclure2011}
{McLure}, R.~J., {et~al.} 2011, ArXiv e-prints (arXiv:1102.4881)

\bibitem[{{McQuinn} {et~al.}(2007){McQuinn}, {Hernquist}, {Zaldarriaga}, \&
  {Dutta}}]{mcquinn2007}
{McQuinn}, M., {Hernquist}, L., {Zaldarriaga}, M., \& {Dutta}, S. 2007, \mnras,
  381, 75

\bibitem[{{Mesinger} \& {Furlanetto}(2008)}]{mesinger2008b}
{Mesinger}, A., \& {Furlanetto}, S.~R. 2008, \mnras, 386, 1990

\bibitem[{{Mesinger} {et~al.}(2004){Mesinger}, {Haiman}, \&
  {Cen}}]{mesinger2004}
{Mesinger}, A., {Haiman}, Z., \& {Cen}, R. 2004, \apj, 613, 23

\bibitem[{{Miralda-Escud{\'e}} {et~al.}(2000){Miralda-Escud{\'e}}, {Haehnelt},
  \& {Rees}}]{miralda2000}
{Miralda-Escud{\'e}}, J., {Haehnelt}, M., \& {Rees}, M.~J. 2000, \apj, 530, 1

\bibitem[{{Nagao} {et~al.}(2006){Nagao}, {Maiolino}, \& {Marconi}}]{nagao2006}
{Nagao}, T., {Maiolino}, R., \& {Marconi}, A. 2006, \aap, 447, 863

\bibitem[{{Nakamura} {et~al.}(2011){Nakamura}, {Inoue}, {Hayashino}, {Horie},
  {Kousai}, {Fujii}, \& {Matsuda}}]{nakamura2011}
{Nakamura}, E., {Inoue}, A.~K., {Hayashino}, T., {Horie}, M., {Kousai}, K.,
  {Fujii}, T., \& {Matsuda}, Y. 2011, \mnras, 412, 2579

\bibitem[{{Oesch} {et~al.}(2010){Oesch}, {Bouwens}, {Illingworth}, {Carollo},
  {Franx}, {Labb{\'e}}, {Magee}, {Stiavelli}, {Trenti}, \& {van
  Dokkum}}]{oesch2010}
{Oesch}, P.~A., {et~al.} 2010, \apjl, 709, L16

\bibitem[{{Oesch} {et~al.}(2011){Oesch}, {Bouwens}, {Illingworth}, {Labbe},
  {Trenti}, {Gonzalez}, {Carollo}, {Franx}, {van Dokkum}, \&
  {Magee}}]{oesch2011}
---. 2011, ArXiv e-prints (arXiv:1105.2297)

\bibitem[{{Oke} \& {Gunn}(1983)}]{oke1983}
{Oke}, J.~B., \& {Gunn}, J.~E. 1983, \apj, 266, 713

\bibitem[{{Ono} {et~al.}(2010){Ono}, {Ouchi}, {Shimasaku}, {Dunlop}, {Farrah},
  {McLure}, \& {Okamura}}]{ono2010}
{Ono}, Y., {Ouchi}, M., {Shimasaku}, K., {Dunlop}, J., {Farrah}, D., {McLure},
  R., \& {Okamura}, S. 2010, \apj, 724, 1524

\bibitem[{{Ouchi} {et~al.}(2004){Ouchi}, {Shimasaku}, {Okamura}, {Furusawa},
  {Kashikawa}, {Ota}, {Doi}, {Hamabe}, {Kimura}, {Komiyama}, {Miyazaki},
  {Miyazaki}, {Nakata}, {Sekiguchi}, {Yagi}, \& {Yasuda}}]{ouchi2004b}
{Ouchi}, M., {et~al.} 2004, \apj, 611, 685

\bibitem[{{Ouchi} {et~al.}(2008){Ouchi}, {Shimasaku}, {Akiyama}, {Simpson},
  {Saito}, {Ueda}, {Furusawa}, {Sekiguchi}, {Yamada}, {Kodama}, {Kashikawa},
  {Okamura}, {Iye}, {Takata}, {Yoshida}, \& {Yoshida}}]{ouchi2008}
---. 2008, \apjs, 176, 301

\bibitem[{{Ouchi} {et~al.}(2009){Ouchi}, {Mobasher}, {Shimasaku}, {Ferguson},
  {Fall}, {Ono}, {Kashikawa}, {Morokuma}, {Nakajima}, {Okamura}, {Dickinson},
  {Giavalisco}, \& {Ohta}}]{ouchi2009b}
---. 2009, \apj, 706, 1136

\bibitem[{{Ouchi} {et~al.}(2010){Ouchi}, {Shimasaku}, {Furusawa}, {Saito},
  {Yoshida}, {Akiyama}, {Ono}, {Yamada}, {Ota}, {Kashikawa}, {Iye}, {Kodama},
  {Okamura}, {Simpson}, \& {Yoshida}}]{ouchi2010}
---. 2010, \apj, 723, 869

\bibitem[{{Pentericci} {et~al.}(2011){Pentericci}, {Fontana}, {Vanzella},
  {Castellano}, {Grazian}, {Dijkstra}, {Boutsia}, {Cristiani}, {Dickinson},
  {Giallongo}, {Giavalisco}, {Maiolino}, {Moorwood}, \&
  {Santini}}]{pentericci2011}
{Pentericci}, L., {et~al.} 2011, ArXiv e-prints (arXiv:1107.1376)

\bibitem[{{Pettini} {et~al.}(2001){Pettini}, {Shapley}, {Steidel}, {Cuby},
  {Dickinson}, {Moorwood}, {Adelberger}, \& {Giavalisco}}]{pettini2001}
{Pettini}, M., {Shapley}, A.~E., {Steidel}, C.~C., {Cuby}, J., {Dickinson}, M.,
  {Moorwood}, A.~F.~M., {Adelberger}, K.~L., \& {Giavalisco}, M. 2001, \apj,
  554, 981

\bibitem[{{Raiter} {et~al.}(2010){Raiter}, {Fosbury}, \&
  {Teimoorinia}}]{raiter2010}
{Raiter}, A., {Fosbury}, R.~A.~E., \& {Teimoorinia}, H. 2010, \aap, 510, A109+

\bibitem[{{Reddy} {et~al.}(2008){Reddy}, {Steidel}, {Pettini}, {Adelberger},
  {Shapley}, {Erb}, \& {Dickinson}}]{reddy2008}
{Reddy}, N.~A., {Steidel}, C.~C., {Pettini}, M., {Adelberger}, K.~L.,
  {Shapley}, A.~E., {Erb}, D.~K., \& {Dickinson}, M. 2008, \apjs, 175, 48

\bibitem[{{Rhoads} {et~al.}(2000){Rhoads}, {Malhotra}, {Dey}, {Stern},
  {Spinrad}, \& {Jannuzi}}]{rhoads2000}
{Rhoads}, J.~E., {Malhotra}, S., {Dey}, A., {Stern}, D., {Spinrad}, H., \&
  {Jannuzi}, B.~T. 2000, \apjl, 545, L85

\bibitem[{{Rhoads} {et~al.}(2003){Rhoads}, {Dey}, {Malhotra}, {Stern},
  {Spinrad}, {Jannuzi}, {Dawson}, {Brown}, \& {Landes}}]{rhoads2003}
{Rhoads}, J.~E., {et~al.} 2003, \aj, 125, 1006

\bibitem[{{Salpeter}(1955)}]{salpeter1955}
{Salpeter}, E.~E. 1955, \apj, 121, 161

\bibitem[{{Santos} {et~al.}(2004){Santos}, {Ellis}, {Kneib}, {Richard}, \&
  {Kuijken}}]{santos2004}
{Santos}, M.~R., {Ellis}, R.~S., {Kneib}, J., {Richard}, J., \& {Kuijken}, K.
  2004, \apj, 606, 683

\bibitem[{{Schaerer} \& {de Barros}(2009)}]{schaerer2009}
{Schaerer}, D., \& {de Barros}, S. 2009, \aap, 502, 423

\bibitem[{{Schaerer} \& {de Barros}(2010)}]{schaerer2010}
---. 2010, \aap, 515, A73+

\bibitem[{{Schenker} {et~al.}(2011){Schenker}, {Stark}, {Ellis}, {Robertson},
  {Dunlop}, {McLure}, {Kneib}, \& {Richard}}]{schenker2011}
{Schenker}, M.~A., {Stark}, D.~P., {Ellis}, R.~S., {Robertson}, B.~E.,
  {Dunlop}, J.~S., {McLure}, R.~J., {Kneib}, J.~., \& {Richard}, J. 2011, ArXiv
  e-prints (arXiv:1107.1261)

\bibitem[{{Shapley} {et~al.}(2003){Shapley}, {Steidel}, {Pettini}, \&
  {Adelberger}}]{shapley2003}
{Shapley}, A.~E., {Steidel}, C.~C., {Pettini}, M., \& {Adelberger}, K.~L. 2003,
  \apj, 588, 65

\bibitem[{{Shim} {et~al.}(2011){Shim}, {Chary}, {Dickinson}, {Lin}, {Spinrad},
  {Stern}, \& {Yan}}]{shim2011}
{Shim}, H., {Chary}, R.-R., {Dickinson}, M., {Lin}, L., {Spinrad}, H., {Stern},
  D., \& {Yan}, C.-H. 2011, \apj, 738, 69

\bibitem[{{Shimasaku} {et~al.}(2006){Shimasaku}, {Kashikawa}, {Doi}, {Ly},
  {Malkan}, {Matsuda}, {Ouchi}, {Hayashino}, {Iye}, {Motohara}, {Murayama},
  {Nagao}, {Ohta}, {Okamura}, {Sasaki}, {Shioya}, \&
  {Taniguchi}}]{shimasaku2006}
{Shimasaku}, K., {et~al.} 2006, \pasj, 58, 313

\bibitem[{{Sokasian} {et~al.}(2002){Sokasian}, {Abel}, \&
  {Hernquist}}]{sokasian2002}
{Sokasian}, A., {Abel}, T., \& {Hernquist}, L. 2002, \mnras, 332, 601

\bibitem[{{Stanway} {et~al.}(2008){Stanway}, {Bremer}, \&
  {Lehnert}}]{stanway2008}
{Stanway}, E.~R., {Bremer}, M.~N., \& {Lehnert}, M.~D. 2008, \mnras, 385, 493

\bibitem[{{Stanway} {et~al.}(2007){Stanway}, {Bunker}, {Glazebrook}, {Abraham},
  {Rhoads}, {Malhotra}, {Crampton}, {Colless}, \& {Chiu}}]{stanway2007}
{Stanway}, E.~R., {et~al.} 2007, \mnras, 376, 727

\bibitem[{{Stark} {et~al.}(2010){Stark}, {Ellis}, {Chiu}, {Ouchi}, \&
  {Bunker}}]{stark2010}
{Stark}, D.~P., {Ellis}, R.~S., {Chiu}, K., {Ouchi}, M., \& {Bunker}, A. 2010,
  \mnras, 408, 1628

\bibitem[{{Stark} {et~al.}(2011){Stark}, {Ellis}, \& {Ouchi}}]{stark2010b}
{Stark}, D.~P., {Ellis}, R.~S., \& {Ouchi}, M. 2011, \apjl, 728, L2+

\bibitem[{{Steidel} {et~al.}(2010){Steidel}, {Erb}, {Shapley}, {Pettini},
  {Reddy}, {Bogosavljevi{\'c}}, {Rudie}, \& {Rakic}}]{steidel2010}
{Steidel}, C.~C., {Erb}, D.~K., {Shapley}, A.~E., {Pettini}, M., {Reddy}, N.,
  {Bogosavljevi{\'c}}, M., {Rudie}, G.~C., \& {Rakic}, O. 2010, \apj, 717, 289

\bibitem[{{Steidel} {et~al.}(1996){Steidel}, {Giavalisco}, {Pettini},
  {Dickinson}, \& {Adelberger}}]{steidel1996a}
{Steidel}, C.~C., {Giavalisco}, M., {Pettini}, M., {Dickinson}, M., \&
  {Adelberger}, K.~L. 1996, \apjl, 462, L17+

\bibitem[{{Stern} {et~al.}(2005){Stern}, {Yost}, {Eckart}, {Harrison},
  {Helfand}, {Djorgovski}, {Malhotra}, \& {Rhoads}}]{stern2005}
{Stern}, D., {Yost}, S.~A., {Eckart}, M.~E., {Harrison}, F.~A., {Helfand},
  D.~J., {Djorgovski}, S.~G., {Malhotra}, S., \& {Rhoads}, J.~E. 2005, \apj,
  619, 12

\bibitem[{{Vanzella} {et~al.}(2009){Vanzella}, {Giavalisco}, {Dickinson},
  {Cristiani}, {Nonino}, {Kuntschner}, {Popesso}, {Rosati}, {Renzini}, {Stern},
  {Cesarsky}, {Ferguson}, \& {Fosbury}}]{vanzella2009}
{Vanzella}, E., {et~al.} 2009, \apj, 695, 1163

\bibitem[{{Vanzella} {et~al.}(2010){Vanzella}, {Grazian}, {Hayes},
  {Pentericci}, {Schaerer}, {Dickinson}, {Cristiani}, {Giavalisco}, {Verhamme},
  {Nonino}, \& {Rosati}}]{vanzella2010}
---. 2010, \aap, 513, A20+

\bibitem[{{Vanzella} {et~al.}(2011){Vanzella}, {Pentericci}, {Fontana},
  {Grazian}, {Castellano}, {Boutsia}, {Cristiani}, {Dickinson}, {Gallozzi},
  {Giallongo}, {Giavalisco}, {Maiolino}, {Moorwood}, {Paris}, \&
  {Santini}}]{vanzella2010d}
---. 2011, \apjl, 730, L35+

\bibitem[{{Wilkins} {et~al.}(2010){Wilkins}, {Bunker}, {Ellis}, {Stark},
  {Stanway}, {Chiu}, {Lorenzoni}, \& {Jarvis}}]{wilkins2010}
{Wilkins}, S.~M., {Bunker}, A.~J., {Ellis}, R.~S., {Stark}, D., {Stanway},
  E.~R., {Chiu}, K., {Lorenzoni}, S., \& {Jarvis}, M.~J. 2010, \mnras, 403, 938

\bibitem[{{Wilkins} {et~al.}(2011){Wilkins}, {Bunker}, {Lorenzoni}, \&
  {Caruana}}]{wilkins2011}
{Wilkins}, S.~M., {Bunker}, A.~J., {Lorenzoni}, S., \& {Caruana}, J. 2011,
  \mnras, 411, 23

\bibitem[{{Yan} {et~al.}(2010){Yan}, {Windhorst}, {Hathi}, {Cohen}, {Ryan},
  {O'Connell}, \& {McCarthy}}]{yan2010}
{Yan}, H., {Windhorst}, R.~A., {Hathi}, N.~P., {Cohen}, S.~H., {Ryan}, R.~E.,
  {O'Connell}, R.~W., \& {McCarthy}, P.~J. 2010, Research in Astronomy and
  Astrophysics, 10, 867

\bibitem[{{Yan} {et~al.}(2011){Yan}, {Yan}, {Zamojski}, {Windhorst},
  {McCarthy}, {Fan}, {R{\"o}ttgering}, {Koekemoer}, {Robertson}, {Dav{\'e}}, \&
  {Cai}}]{yan2010b}
{Yan}, H., {et~al.} 2011, \apjl, 728, L22+

\end{thebibliography}

\end{document}